\documentclass[12pt,eqsecnum,tightenlines,floats,aps,amsmath,
amssymb,nofootinbib,prd,noshowpacs,notitlepage,longbibliography]{revtex4}
\usepackage[utf8]{inputenc}
\usepackage{amsmath,amsthm,latexsym,amssymb,amsfonts,color,graphicx}
\usepackage{hyperref,epstopdf}

\global\arraycolsep=1truept
\newcommand{\beq}{\begin{equation}}
\newcommand{\eeq}{\end{equation}}
\def\bea{\begin{eqnarray}}
\def\eea{\end{eqnarray}}
\definecolor{darkred}{rgb}{.8,0,0}

\definecolor{darkblu}{rgb}{0,0,.8}

\def\vs{\vspace{4mm}}

\newcommand{\BOX}[1]{
\begin{center}\fbox{\parbox{135mm}{#1}}\end{center}}

\newcommand{\BOXthree}[1]{
\begin{center}\fbox{\parbox{140mm}{#1}}\end{center}}
\newcommand{\BOXfour}[1]{
\begin{center}\fbox{\parbox{105mm}{#1}}\end{center}}
\newcommand{\BOXfive}[1]{
\begin{center}\fbox{\parbox{125mm}{#1}}\end{center}}
\newcommand{\BOXsix}[1]{
\begin{center}\fbox{\parbox{150mm}{#1}}\end{center}}

\begin{document}

\title[]{Observers with constant proper acceleration, constant proper jerk, and beyond}

\author{Josep M. Pons{$^{a} $}}\email{pons@fqa.ub.edu}
\author{Ferran de Palol{$^{b} $}}\email{fdepalco@gmail.com}
\affiliation{$^{a}$ Departament de F\'\i sica Qu\`antica i Astrof\'\i{s}ica and
Institut de Ci\`encies del Cosmos (ICCUB), Facultat de F\'\i sica, Universitat de Barcelona, 
Mart\'{\i} Franqu\`es 1,
08028 Barcelona, Catalonia, Spain.}
\affiliation{$^{b} $ Facultat de F\'\i sica, Universitat de Barcelona, 
Mart\'{\i} Franqu\`es 1,
08028 Barcelona, Catalonia, Spain.}

\vskip 2truecm


\bigskip
\begin{abstract}

We discuss in Minkowski spacetime the differences between the concepts of constant proper 
$n$-acceleration and of vanishing 
$(n+1)$-acceleration. By $n$-acceleration we essentially mean the higher order time derivatives of the 
position vector of the trajectory of a point particle, adapted to Minkowski spacetime or eventually to 
curved spacetime. The $2$-acceleration is known as the Jerk, the $3$-acceleration as the Snap, etc. 
As for the concept of {\sl proper} $n$-acceleration we give a specific definition involving 
the instantaneous comoving frame of the observer and we discuss, in such framework, 
the difficulties in finding a characterization of this notion as a Lorentz invariant statement. 
We show how the Frenet-Serret formalism helps to address the problem. 
In particular we find that our definition of an observer 
with constant proper acceleration corresponds to the vanishing of the third curvature invariant 
$\kappa_3$ (thus the motion is three dimensional in Minkowski spacetime) together with the constancy 
of the first and second curvature invariants and the restriction $\kappa_2 < \kappa_1$, 
the particular case $\kappa_2=0$ being the one commonly referred to in the literature.
We generalize these concepts to curved spacetime, in which the notion of trajectory in a plane is 
replaced by the vanishing of the second curvature invariant $\kappa_2$. Under this condition, the 
concept of constant proper $n$-acceleration coincides with that of the vanising of the $(n+1)$-acceleration 
and is characterized by the fact that the first curvature invariant $\kappa_1$ is a 
$(n-1)$-degree polynomial of proper time. We  illustrate some of our results with examples in Minkowski,
de Sitter and Schwarzschild spacetimes.
\end{abstract}


\maketitle

\vskip 1truecm
\vfill\eject

\section{Introduction}
\label{intro}

Although the physics in Minkowski spacetime as described by an inertial observer 
is indeed a very relevant description (one can just take a look at quantum
field theory textbooks), the use of non inertial frames of reference,
like the ones associated with accelerated observers,
has become a very rich territory for our present understanding of
new physical phenomena emerging from the quantum world. This is the case for instance of
the thermal bath of Unruh radiation \cite{Unruh:1976db}
that a constantly accelerated observer\footnote{From now on, by constantly 
accelerated observer we mean the observer with constant {\sl proper} 
acceleration. Proper in the sense of being described in the instantaneous frame comoving with the observer.} 
should experience, though it has not been experimentally verified yet
\cite{Crispino:2007eb} \cite{Cozzella:2018byy}. 
In addition, and just by taking into account the equivalence principle, 
one can relate the physics experienced by these accelerated observers 
in Minkowski spacetime 
with that of an observer hovering above the horizon of a Schwarzschild black hole. 
In fact the local description of a Schwarzschild black 
hole near the horizon is that of a Rindler spacetime, which is nothing but a patch of Minkowski 
spacetime described with the coordinates which are natural to the constantly accelerated observer
(the coordinates in which this observer remains at rest).

Beyond the concept of acceleration there is an infinite tower of 
kinematical concepts with higher order time derivatives, which go under the names of Jerk, 
Snap, Crackle, etc.\footnote{We learn in \cite{Dunajski:2008tg} that ``This terminology
goes back to a 1932 advertisement of Kellogg's Rice Crispies which 'merrily Snap, crackle, and 
pop in a bowl of milk' ''. Here our use of these concepts is unrelated to the standard use in 
cosmology as higher order time derivatives of the scale factor in the 
Friedmann-Lemaitre-Robertson-Walker (FLRW) metric}. 
In Galilean mechanics these objects are trivial generalizations of the 
concept of acceleration, and their extension to special relativity has been 
introduced in \cite{Russo:2009yd} in a way that can be immediately generalized to curved spacetimes.

The main purpose of this paper is to elucidate the differences between the concepts of 
constant proper $n$-acceleration and vanishing $(n+1)$-acceleration. It is crucial in this respect
to properly define the instantaneous comoving frame. We choose to do it in
the simplest way available by implementing the map form the original frame to the  
instantaneous comoving frame as a pure boost, wih no additional rotations attached.

\vs

The organization of the paper is as follows. In section \ref{accobs} we derive in a 
non conventional way the equations for the trajectories of the observer with constant proper
acceleration in 4d (or any dimension) Minkowski spacetime. From our analysis we notice that
the trajectory of the observer with constant proper acceleration can be three dimensional (3d)
or two-dimensional (2d). The 3d case, which is new, is solved in section \ref{the3dsol}.
The notion of Jerk is introduced in section \ref{EnterstheJerk} and it is shown that its   
vanishing characterizes the motion of the observer
with constant proper acceleration in the 2d case, that is, when this motion takes place in a plane.

In the next section \ref{JerkSnapCrackle} we introduce all the tower of kinematical concepts with 
higher order time derivatives and in section \ref{vanishingSnap} we address in detail the 
case of vanishing Snap and show that, unlike what would have been analogous to the vanishing 
Jerk case, it does not imply that 
the motion has constant proper Jerk, nor that the motion is restricted to a plane. 
We further deal with this issue in section \ref{CpAJS} in which 
we show the difficulties of finding a Lorentz invariant statement for notions like that of a constant 
proper acceleration or a constant proper Jerk, etc. 
The Frenet-Serret formalism is introduced in section \ref{FS} as a possible way out of these 
difficulties. In this way we are able to characterize the 3d motion of the observer 
with constant proper acceleration (section \ref{revisit}). We obtain in section \ref{2d-solved} 
solutions for the trajectory of an observer with vanishing $n$-acceleration and constant 
$(n-1)$-acceleration in a 2d Minkowski spacetime. Conclusions and discussion are given 
in section \ref{concl}.

We finish with three appendices. The first is devoted to the circular motion in Minkowski spacetime; 
the second to constant proper acceleration in de Sitter spacetime; and the third to
circular orbits in Schwarzschild spcetime.

\vspace{4mm}

\section{The accelerated observer in 4d Minkowski spacetime}
\label{accobs}

Given the trajectory $x^{\mu}(t(s))$ with $t$ the coordinate time and $s$ the proper time, 
its velocity with respect to the proper time is
$$V(s)=\gamma\left(
\begin{array}{c}
1\\x'(t(s))\\y'(t(s))\\z'(t(s))\\
\end{array}
\right) = \gamma\left(\begin{array}{c}
1\\ \vec r'(t(s))\\
\end{array}
\right)
$$
where $x'(t)$ is the derivative of $x(t)$ with respect to the coordinate time so that 
$\displaystyle \frac{d x^i}{d s} = \gamma\, x'(t)  
$ with $\displaystyle\gamma = \frac{d t}{d s} =\frac{1}{\sqrt{1-(r')^2}}$ 
(we use dot, $\cdot$, for proper 
time derivative $\displaystyle\frac{d }{d s}$ and prime, $'$, for coordinate time derivative 
$\displaystyle\frac{d }{d t}$)\,. The definition of proper time guarantees $(V)^2=-1$.

\vs

We can prepare a family of Lorentz transformations $B(s)$ so that at any proper time $s$ 
we send the traveler to rest,
$$V(s) \to B(s)\, V(s) = \left(
\begin{array}{c}
1\\0\\0\\0\\
\end{array}
\right) =:V_p\,,\ \forall s\,.
$$
There is an arbitrariness in choosing this family, because any boost sending the observer to rest can
be followed by an arbitrary rotation without affecting the status of the observer as being at rest. 
Our choice will be the most simple available: that of a pure boost. Some of the results and definitions in
this paper depend upon this choice. Other choices can be explored, for instance adding a rotation 
depending on the acceleration, but we think that ours is the first option to be considered.
The pure boost that does the job is (with $t=t(s)$),
\beq
B(s)=\left(
\begin{array}{cccc}
 \gamma  & -\gamma\,  x'(t) & -\gamma\,  y'(t) & -\gamma\,  z'(t) \\
 -\gamma\,  x'(t)\ & 1+\frac{\gamma ^2 }{\gamma +1}x'(t)^2\ & \frac{\gamma ^2 }{\gamma +1}x'(t) y'(t)\ 
 & \frac{\gamma ^2 }{\gamma +1}x'(t)  z'(t) \\
 -\gamma\,  y'(t)\ & \frac{\gamma ^2 }{\gamma +1}x'(t) y'(t)\ & 1+\frac{\gamma ^2 }{\gamma +1}y'(t)^2\ 
 & \frac{\gamma ^2 }{\gamma +1}y'(t)  z'(t) \\
 -\gamma\,  z'(t)\ & \frac{\gamma ^2 }{\gamma +1}x'(t) z'(t)\ & \frac{\gamma ^2 }{\gamma +1}y'(t) z'(t)\ 
 & 1+\frac{\gamma ^2  }{\gamma +1}z'(t)^2\\
\end{array}
\right)\,.
\label{theboost}
\eeq
The existence of non vanishing proper acceleration will be detected if, to do the same job, 
the boost necessary 
at proper time $s+\delta s$ is different from $B(s)$. In such case we will need an additional 
infinitesimal boost $\bar B(\delta s)$ so that $\bar B(\delta s)\,B(s)= B(s+\delta s)$. Thus,
expanding to first order,
$\bar B(\delta s) =(B(s)+\dot B(s)\delta s)\,B^{-1}(s) = I + \dot B(s) B^{-1}(s)\delta s$. 
The acceleration caused by $\bar B(\delta s)$ is, by definition,
$$
\frac{1}{\delta s}\Big(\bar B(\delta s)V_p -V_p\Big) = \dot B(s) B^{-1}(s)V_p\,.
$$
Thus the proper 
acceleration $A_p(s)$ (only spatial components) experienced by the observer at proper time $s$
is given by the action of $-\dot B(s) B^{-1}(s)$ on the rest velocity $V_p$ 
(a minus sign because the infinitesimal boost compensates for the acceleration at proper time $s$). 
Using $\displaystyle\frac{d }{d s} = \gamma\frac{d }{d t}$ we get, 
\beq
A_p(s) = -\dot B(s) B^{-1}(s) V_p= \left(
\begin{array}{c}
0\\a_x\\a_y\\a_z\\
\end{array}
\right)  = \left(
\begin{array}{c}
0\\\vec a\\
\end{array}
\right) \,,
\label{propA}
\eeq
with 
\beq\vec a= \gamma^2 \vec r''+\frac{\gamma }{\gamma+1} \gamma' \vec r'  ,\quad 
\gamma' = \frac{d \gamma}{d t} =\frac{1}{2} \gamma^3 \frac{d }{d t} (\vec r')^2\,.
\label{propA2}
\eeq
Given the proper acceleration $A_p(s)$ in the instantaneous comoving frame, 
we can move back to the original 
coordinate system by applying the inverse boost $B^{-1}(s)$. Thus, 
taking into account that $V_p$ is constant, we obtain
\bea
A(s) &=& B^{-1}(s)A_p(s) = -B^{-1}(s)\dot B(s) B^{-1}(s) V_p=  
\Big(\frac{d }{d s} B^{-1}(s)\Big)\,V_p\nonumber \\ &=& 
\frac{d }{d s}\Big( B^{-1}(s)\,V_p\Big) = \frac{d }{d s}V(s)\,,
\label{asexpected}
\eea
that is, $A(s)=\dot V(s)$, as it could have been expected. 
Again, using $\displaystyle\frac{d }{d s} = \gamma\frac{d }{d t}$, one can 
also write 
\bea
A(s) = \left(
\begin{array}{c}
0\\ \gamma^2 \vec r''   \\
\end{array}
\right)+\gamma'\,V(s) = \left(
\begin{array}{c}
\gamma\gamma'\\ \vec a + \frac{\gamma^2}{\gamma+1}\gamma' \vec r' \\
\end{array}
\right)\,,
\label{vec a}
\eea
and (note that $\gamma' = \vec a \cdot \vec r'$)
$$ A^2 = A_p^2 = \vec a^2 = \gamma^4 (\vec r'')^2 + (\gamma')^2.
$$
\subsection{\bf Constant proper acceleration}

Let us examine the case of constant proper acceleration, $\vec a= \vec a_{\!{}_0}\ constant $. In our view
this constancy will correspond to Einstein's intuitive definition of uniform acceleration 
\cite{Friedman:2016tye}.

The equation of motion in the original coordinate system is, from (\ref{propA2}),
\beq
\vec r''= \frac{1}{\gamma^2}\vec a_{\!{}_0} -\frac{1 }{\gamma(\gamma+1)} \gamma' \vec r'\,.
\label{eqpropacc}
\eeq
This equation is more easily managed with the use of proper time $s$. 
Equation (\ref{eqpropacc}) then becomes (note that $\dot t = \gamma$)
\beq
\ddot{\vec r}(s) = \vec a_{\!{}_0} +\frac{\ddot t (s)}{\dot t(s) +1}\, \dot{ \vec r}(s)\,.
\label{eqpropaccproptime}
\eeq
A parametrization of the velocity which guarantees that $\dot t(s)  >0\,,\ \forall s$, is 
the following,
\beq
V(s) = \left(
\begin{array}{c}
\dot t(s)\\ \dot{\vec  r}(s)  \\
\end{array}
\right) = \left(
\begin{array}{c}
\cosh(f(s))\\ \sinh(f(s))\, \vec n(s) \\
\end{array}
\right) 
\label{Vel}
\eeq
with $\vec n(s)$ a unitary euclidean $3$-vector. We obtain, from
(\ref{eqpropaccproptime})
\beq
\dot {\vec n}= \frac{1 }{\sinh(f)}\Big(\vec a_{\!{}_0} - \dot f \,\vec n  \Big)\,.
\label{dotn}
\eeq
An important consequence of this equation (\ref{dotn}) is that it shows that the motion of 
$\vec n(s)$ in 
$3$-space is at most in a plane containing $\vec a_{\!{}_0}$. From $(\dot {\vec n},{\vec n})=0$ 
we infer that $\dot f =(\vec a_{\!{}_0}, {\vec n})$. Inserting (\ref{dotn}) into its derivative  
$\ddot f =(\vec a_{\!{}_0}, \dot {\vec n})$ we obtain the equation for $f$,
\beq
\ddot f\sinh(f) = a_{\!{}_0}^2-(\dot f)^2\,.
\label{eqf}
\eeq
An obvious solution is $f(s)= a_{\!{}_0}\,(s-s_{\!{}_0})$ with arbitrary $s_{\!{}_0}$ and 
$a_{\!{}_0}= |\vec a_{\!{}_0}|$. 
In this case  $\dot {\vec n}=0$ and 
$\displaystyle \vec n = \frac{\vec  a_{\!{}_0}}{ a_{\!{}_0}}$ and the motion takes place 
in a plane (2d) in Minkowski spacetime. For any other solution, having seen from (\ref{dotn}) 
that the motion of $\vec n(s)$ in $3$-space is in a plane, we infer that the trajectory is 3d in 
Minkowski spacetime.

\subsection{\bf The 2d solution}
Let us focus now on this 2d solution. 
The trajectory can be easily obtained by integrating (\ref{Vel}) with 
constant ${\vec n}$ and $f(s)= a_{\!{}_0}\,(s-s_{\!{}_0})$. One can also work directly with 
(\ref{eqpropacc}) and write $\vec r'(t) = \lambda'(t)\, \vec a_{\!{}_0}$,
for some function $\lambda'(t)$, This equation integrates to 
$\vec r(t) = \vec r_{\!{}_0} +(\lambda(t)-\lambda_{\!{}_0})\, \vec a_{\!{}_0}$\,, 
with $\vec r_{\!{}_0} = \vec r(t_{\!{}_0})$\,, $\lambda_{\!{}_0} = \lambda(t_{\!{}_0})$ for an initial time
$t_{\!{}_0}$.  
Substituting it in (\ref{eqpropacc}) we get for $\lambda(t)$ 
the equation (with $a_{\!{}_0}=|\vec a_{\!{}_0}|$)
$$
\frac{\lambda''}{(1-(\lambda')^2 a_{\!{}_0}^2)^{\frac{3}{2}}} = 1\,,
$$
with general solution
\beq
a_{\!{}_0}^2 (\lambda-\lambda_{\!{}_0})^2 - (t-t_{\!{}_0})^2 =\frac{1}{a_{\!{}_0}^2}\quad  \Rightarrow  \quad
(\vec r-\vec r_{\!{}_0})^2 - (t-t_{\!{}_0})^2 =\frac{1}{a_{\!{}_0}^2}\,,
\label{eqacc}
\eeq
which are the expected equations for the 2d constantly accelerated oberver.

\subsection{\bf The 3d solution}
\label{the3dsol}
To find the 3d solution of the constantly accelerated observer it proves useful 
to adopt a parametrization different from that of (\ref{Vel}). We only need to consider a 3d Minkowski 
spacetime, and we parametrize the velocity (with respect to proper time $s$) of the trajectory with
\beq
V = \left(
\begin{array}{c}
V_0 \\ \vec V \\
\end{array}
\right)\,\quad {\rm with}\quad \vec V =\left(
\begin{array}{c}
V_1 \\ V_2 \\
\end{array}
\right)= v(s) \left(
\begin{array}{c}
\sin(f(s)) \\ \cos(f(s)) \\
\end{array}
\right)\,,
\label{Vfor3d}
\eeq
with $v(s)$ and $f(s)$ to be determined by the conditions of constant proper acceleration. 
To guarantee 
that $(V)^2=-1$ notice that $V_0$ gets determined as  $V_0=\sqrt{v^2+1}$.

The pure boost of eq.(\ref{theboost}), now with a different notation, can be written as 
\beq B
=
\left(
\begin{array}{cc}
\, V_0 \quad & -\vec V  \\
 -\vec V \quad\ & I+\frac{1}{V_0 +1}\vec V \otimes \vec V   \\
\end{array}
\right)\,.
\label{theboost2d3}
\eeq
The action of $B(s)$ on the acceleration $\displaystyle A=\frac{d}{d\,s}V$ defines the 
proper acceleration $A_p$\,,
$$
B\,A=A_p= \left(
\begin{array}{c}
0 \\ \vec A_p \\
\end{array}
\right)\,,
$$
and now $\vec A_p$ is required to be a constant vector $\vec a_{\!{}_0}$.
Using (\ref{Vfor3d}), (\ref{theboost2d3}), we find
$$
\vec A_p= \frac{\dot v}{V_0}\left(
\begin{array}{c}
\sin(f(s)) \\ \cos(f(s)) \\
\end{array}
\right) +  v\dot f\left(
\begin{array}{c}
\cos(f(s)) \\ -\sin(f(s)) \\
\end{array}
\right) = \vec a_{\!{}_0}= a_{\!{}_0} \left(
\begin{array}{c}
\cos\theta_{\!{}_0} \\ \sin\theta_{\!{}_0} \\
\end{array}
\right)\,,
$$
with constant $\theta_{\!{}_0}$. From this expression we infer 
\beq\frac{\dot v}{V_0} = a_{\!{}_0} \sin(f(s)+\theta_{\!{}_0})\,,\quad v\dot f = a_{\!{}_0} \cos(f(s)+\theta_{\!{}_0})\,,
\label{the3dsol1}
\eeq
and, consequently,
$$ \frac{\dot v}{V_0}\cos(f(s)+\theta_{\!{}_0}) = v\dot f\sin(f(s)+\theta_{\!{}_0}) = 
- v \frac{d}{d\,s}\cos(f(s)+\theta_{\!{}_0})\,,
$$
or, defining $g(s)= \cos(f(s)+\theta_{\!{}_0})$\,,
$$ \frac{\dot g}{g}= -\frac{\dot v}{v\,V_0}\,,
$$
which integrates to ($V_0=\sqrt{v^2+1}$)
\beq
g\, \frac{v}{1 +V_0} = c\,,
\label{theg}
\eeq
with $c$ a constant that must be $|c|<1$ because $|g|\leq 1$ and $|\frac{v}{1 +V_0}|<1$. Thus
$$
\cos(f(s)+\theta_{\!{}_0})=g(s)= \frac{c\,(1+V_0)}{v}\,,
$$
which can be inserted into the first equation in (\ref{the3dsol1}),
$$
\frac{\dot v}{V_0} = a_{\!{}_0} \sqrt{1-g^2} =  a_{\!{}_0} \sqrt{1-\frac{c^2(1+V_0)^2}{v^2}} \,,
$$
Thus,
$$\dot V_0 = \frac{v\,\dot v}{V_0} = a_{\!{}_0} \sqrt{v^2-c^2(1+V_0)^2} = 
a_{\!{}_0} \sqrt{(1+V_0)(d (1+V_0)-2)}\,
$$
with $d = 1-c^2$ which implies $0<d\leq 1$.
Finally, defining the new variable $z(s)$ from $1+V_0 =\frac{1}{d}(z+1)$, 
the last equation becomes
$$
\dot z = \sqrt{d}\, a_{\!{}_0} \sqrt{z^2-1}\ \Rightarrow \ 
z(s) = \cosh(\sqrt{d}\,  a_{\!{}_0} (s-s_{\!{}_0}))\,.
$$
With the choice $\theta_{\!{}_0}=0$, which is always reachable by rotating the spatial coordinates, 
we obtain, using (\ref{theg}),
\bea \displaystyle V_1&=& v \sin(f(s)) = v \sqrt{1-g^2} = \sqrt{(1+V_0)(d (1+V_0)-2)}= 
\frac{1}{\sqrt d}\sqrt{z^2-1}\nonumber\\  
&=& \frac{1}{\sqrt d}\sinh(\sqrt{d}\,  a_{\!{}_0} (s-s_{\!{}_0}))\,,\nonumber\\ 
V_2 &=& v \cos(f(s)) = v \, g = c\,(1+V_0) =  \frac{\sqrt{1-d}}{d}(z+1)\nonumber\\ 
&=& \frac{\sqrt{1-d}}{d} (\cosh(\sqrt{d}\,  a_{\!{}_0} (s-s_{\!{}_0})) +1)\nonumber\,.
\eea
Thus the solution for $V$ is
\bea
V &=& \left(
\begin{array}{c}
V_0 \\ \vec V \\
\end{array}
\right) = \left(
\begin{array}{c}
\frac{1}{d} (\cosh(\sqrt{d}\,  a_{\!{}_0} (s-s_{\!{}_0})) +1)-1
\\ \frac{1}{\sqrt d}\sinh(\sqrt{d}\,  a_{\!{}_0} (s-s_{\!{}_0})) \\  
\frac{\sqrt{1-d}}{d} (\cosh(\sqrt{d}\,  a_{\!{}_0} (s-s_{\!{}_0})) +1)\\
\end{array}
\right)\,
\label{V-3d}
\eea
and for the position vector $X$,
\bea
X(s)&=&\left(
\begin{array}{c}
t \\ x \\ y\\ 
\end{array}
\right)= X(s_{\!{}_0}) + \left(
\begin{array}{c}
\frac{1}{a_{\!{}_0}\, d ^{3/2}}\sinh \left(a_{\!{}_0} \sqrt{d } (s-s_{\!{}_0})\right)
+\left(\frac{1-d}{d }\right) (s-s_{\!{}_0})
\\ \frac{1}{a_{\!{}_0} d }\left(\cosh \left(a_{\!{}_0} \sqrt{d } (s-s_{\!{}_0})\right)-1\right)
\\  \frac{\sqrt{1-d } }{a_{\!{}_0} d ^{3/2}}\left(a_{\!{}_0} \sqrt{d } (s-s_{\!{}_0})
+\sinh \left(a_{\!{}_0} \sqrt{d } (s-s_{\!{}_0})\right)\right)
\\ 
\end{array}
\right)
\label{XV-3d}
\eea
The motion in $2$-space, coordinates $(x,y)$, exhibits asymptotes.
Note that $d$ is related to the angle $\delta$ (the ``scattering angle'') 
in the $(x,y)$ plane between the ingoing 
$s\to -\infty$ and the outgoing $s\to \infty$ directions, 
$\displaystyle d= (\sin\frac{\delta}{2})^2$. For 
$d=1\, (\delta=\pi)$ we recover the 2d solution described in (\ref{eqacc}), 
with a turning point for the trajectory 
in $1$-space.  
$$
X_{(d=1)}(s) = \frac{1}{a_{\!{}_0}}\left(
\begin{array}{c}
 \sinh( a_{\!{}_0} (s-s_{\!{}_0})) 
\\ \cosh(  a_{\!{}_0} (s-s_{\!{}_0}))-1 \\  
0\\
\end{array}
\right)\,,
$$

\vs
 
We will revisit this solution in section \ref{revisit}.


\subsection{Enters the Jerk}
\label{EnterstheJerk}
As anticipated above, we will see in section \ref{CpAJS} that the concept of constant proper 
acceleration, as defined by way of (\ref{propA}) with constant $\vec a$, depends on the coordinate frame 
used\footnote{This issue only arises if the 
Minkowski spacetime has more than $2$ dimensions.}, and this is not completely satisfactory.
One should really say that a trajectory
exhibits constant proper acceleration in Minkowski spacetime if {\sl there exist} a coordinate frame 
for which the object $\vec a$ 
defined in  (\ref{propA2}) is constant. There is nothing wrong with such definition but it is much more 
convenient and elegant to have a definition in the form of an invariant statement -a covariant equation 
which in addition we wish to carry over to general relativity. 
In the case of the 2d motion, 
this is achieved with the notion of the Jerk.

The Jerk vector $\Sigma$ (we follow the notation of \cite{Russo:2009yd}) is defined as the 
component orthogonal to the velocity of the proper time derivative of the acceleration 
(see section \ref{JerkSnapCrackle} for more details),
\beq
\Sigma = \frac{d\,A}{d\,s} - (A)^2\,V\,, 
\label{defSigma}
\eeq
Now it is easy to show that for an observer with constant proper acceleration and 2d motion
the Jerk vector vanishes -which is an invariant statement. 
In fact $\Sigma$ is orthogonal to $V$ by construction (see next section \ref{JerkSnapCrackle}
for details) and if the proper acceleration is constant then $\Sigma$ is also orthogonal to $A$ because 
-considering (\ref{defSigma})- $A$ is orthogonal 
to $V$ and has $(A)^2\equiv (A,\,A)$ constant. Then if the motion is in a plane and $\Sigma$ 
is orthogonal both to $V$ and $A$, it must vanish. 

On the other hand, the vanishing of the Jerk leads to the constantly accelerated trajectories in 
2d motion examined above, 
as we now show. $\Sigma =0$ implies $\displaystyle \dot A = (A)^2\,V$, so we have 
$(A.\dot A)=0$, which means $(A)^2$ is constant, $(A)^2=a_{\!{}_0}^2$ with $a_{\!{}_0}=|A|$. Then we have 
(defining the normalized vector $\displaystyle A_N:= \frac{A}{|A|}$) the equations,
\bea
\dot V&=&A =: a_{\!{}_0} A_N \nonumber\\
\dot A_N&=&\frac{1}{a_{\!{}_0}}\dot A=  a_{\!{}_0}\,V\,,
\label{ctproperA}
\eea
which show that the solution is indeed 2d. The solution can be written as
\beq
V(s)= \cosh(a_{\!{}_0} s) V(0) + \sinh(a_{\!{}_0} s) A_N(0) 
\label{2d-ct-acc-obs}
\eeq
where the initial conditions are restricted to satisfy $V(0)^2= -1,\ A_N(0)^2=1$ and 

\noindent $(V(0)\cdot A_N(0))=0$. 

The trajectory is
$$
X(s)= X(0) + \frac{1}{a_{\!{}_0}} \Big(\sinh(a_{\!{}_0} s) V(0) + (\cosh(a_{\!{}_0} s) -1) A_N(0) \Big)\,,
$$
that is, the well known constantly accelerated observer.

\vs

If we parametrize the initial conditions in a 2d subspace of Minkowski,
$$V(0) = \left(
\begin{array}{c}
\cosh(a_{\!{}_0}  s_{\!{}_0}) \\-\sinh(a_{\!{}_0}  s_{\!{}_0})\\
\end{array}
\right)\, 
\qquad 
A_N(0) = \left(
\begin{array}{c}
-\sinh(a_{\!{}_0}  s_{\!{}_0})\\ \cosh(a_{\!{}_0}  s_{\!{}_0})\\
\end{array}
\right)\,,
$$
for some real $s_{\!{}_0}$, we may write
$$V(s) = \left(
\begin{array}{c}
\cosh(a_{\!{}_0}( s-s_{\!{}_0})) \\\sinh(a_{\!{}_0}( s-s_{\!{}_0}) )\\
\end{array}
\right)\,,
$$
which is nothing but (\ref{Vel}) with $f(s)= a_{\!{}_0}( s-s_{\!{}_0})$.

We conclude that, in Minkowski spacetime -no matter the number of dimensions-, if the motion 
takes place in a plane (2d), 
the observer with constant proper acceleration is identified by the vanishing of the Jerk, $\Sigma =0$.

\BOXfive{\vfill\vspace{1.4mm}
\it In Minkowski spacetime: 

Jerk $=0$\ $\Leftrightarrow$\   Constant proper acceleration\ $+$\ Two-dimensional  motion.

\vfill\vspace{1.5mm}}
\label{eqaccBOX}
\subsection{Extension to curved spacetime}
\label{Extension}
The important novelty is that this statement on the vanishing of the Jerk can be exported to
curved spacetime. In fact, the definition of the Jerk (and
all the tower of kinematical concepts) is extended to curved spacetime with the replacement of the 
 partial derivative by the covariant derivative,
$$
\partial_\mu \rightarrow \nabla_\mu\qquad \Rightarrow \qquad \frac{d}{d\,s}A^\mu \rightarrow
\frac{d}{d\,s}A^\mu + V^\rho \Gamma_{\rho\nu}^\mu A^\nu\,,
$$
and requiring the connection to satisfy the metricity condition. The reason for this is the following. 
Let us define $\dot A$ in curved spacetime as the geometric object displayed just above, that is,
$\displaystyle \dot A = \frac{d}{d\,s}A + V \Gamma A$, then, in order to 
guarantee that results like $(\dot A ,V) = -(A)^2$, or 
$\displaystyle \frac{d}{d\,s} (A,A) = 2 ( \dot A, A)$,
still hold in curved spacetime, we need the 
covariant derivative of the metric to vanish. With this condition the Lorentz covariant expressions
in this section and the next sections can be exported to curved spacetime.
Since the Levi Civita connection is metric compatibile, 
we conclude that in the framework of general 
relativity the vanishing of the Jerk ensures that the observer undergoes constant proper 
acceleration. This is also  
true for other generally covariant theories with metric compatibile connections \cite{Hehl:1994ue}, 
like the Einstein-Cartan theory, in which torsion is sourced by fermionic matter, 
or teleparallel gravity, defined with the Weitzenb\"ock connection.

\BOXsix{\vfill\vspace{1.4mm} 
\it In curved spacetime: 

The vanishing of the Jerk implies that the observer has constant proper acceleration. 
\vfill\vspace{1.5mm}}
\label{eqaccBOXcurved}

We give an example in Appendix \ref{B} of constant proper acceleration in de Sitter spacetime. 

\section{Jerk, Snap, Crackle...}
\label{JerkSnapCrackle}
Acceleration, Jerk, Snap, Crackle, Pop, etc., are a family of kinematical concepts that are 
built with the second, third, fourth, etc, derivatives -with respect to proper time- of the 
trajectory (or with the first, second, third derivatives of the velocity 
vector). Each one is defined (see \cite{Russo:2009yd}) 
as the component orthogonal to the velocity of the proper time derivative of the previous object. This 
definition guarantees a threefold condition, namely, that all these objects are spacelike, 
that in the instantaneous comoving frame of the observer, they have only
spatial components, as it already happens with the acceleration, and that if one of them vanishes, 
say the Jerk, all the higher derivative objects, say the Snap, Crackle, etc, will vanish as well. 

Thus acceleration $A$, defined as (see (\ref{asexpected}))
\beq A= \frac{d}{d s} V \,, 
\label{theA}
\eeq
is already orthogonal to the velocity, since $(V)^2 = (V , V) = -1$\,. The Jerk $ \Sigma$ is then defined as 
\beq \Sigma = \left(\frac{d}{d s} A\right)_{\!\perp}= \frac{d}{d s} A -\frac{1}{(V)^2} (\frac{d}{d s} A, V)\, V = \frac{d}{d s} A + 
(\frac{d}{d s} A, V)\, V =\frac{d}{d s} A - (A)^2 V\,,
\label{theJ}
\eeq
and the Snap $\Xi$,
\beq 
\Xi = \left(\frac{d}{d s} \Sigma\right)_{\!\perp}= 
\frac{d}{d s} \Sigma - \frac{1}{(V)^2}(\frac{d}{d s} \Sigma, V) V= \frac{d}{d s} \Sigma + (\frac{d}{d s} \Sigma, V) V 
= \frac{d}{d s} \Sigma - (\Sigma,A) V\,,
\label{theS}
\eeq
and so on. 

In order to generalize these expressions above, 
we adopt the notation of $A_1$ for the acceleration $A$; $A_2$ for 
the ``second acceleration'' or Jerk, 
$A_2=\Sigma$; $A_3$ for the  ``third acceleration'' or Snap, $A_3=\Xi$, etc. In general the $n$-acceleration
will be denoted by $A_n$. The recursive definition of these objects is then
\beq 
 A_1= \frac{d}{d s} V\,,\qquad A_{(n+1)} = \left(\frac{d}{d s} A_n\right)_{\!\perp}= 
 \frac{d}{d s} A_n -( A_n,A_1) V \,.
\label{recursive}
\eeq
Having the case of vanishing Jerk been discussed in the previous section, 
we shall address next the case of vanishing Snap.

\section{Vanishing Snap}
\label{vanishingSnap}
Let us discuss the case $ \Xi =0$. This implies $\displaystyle\frac{d}{d s} \Sigma = (\Sigma,A) V$ and 
therefore 
$\displaystyle\frac{d}{d s}(\Sigma^2) =  2(\Sigma,\dot\Sigma) = 0$, which means that $\Sigma^2$ is constant, 
$\Sigma^2 = \sigma^2$ with $\sigma$ real -and we take $\sigma>0$ (the $\sigma=0$ case corresponds to the 
vanishing Jerk case discussed in the previous section). Note also that 
$\displaystyle \frac{d}{d s}(\Sigma,A) = \Sigma^2= \sigma^2$ and therefore $(\Sigma,A) = \sigma^2 (s - s_{\!{}_0})$ 
with the arbitrary real constant $s_{\!{}_0}$. Now we can compute 
$\displaystyle\frac{d}{d s} (A^2) = 2 (\Sigma,A) = 2 \sigma^2 (s - s_{\!{}_0}) $ and we get 
$A^2 = \sigma^2 (s - s_{\!{}_0})^2 + \beta^2$, where we have implemented the condition of $A^2 \geq 0$ 
with $\beta$ real\,.
 We end up with
\bea A^2 &=& \sigma^2 (s - s_{\!{}_0})^2 + \beta^2\,, \nonumber\\
(\Sigma,A) &=& \sigma^2 (s - s_{\!{}_0})\,,\nonumber \\
\Sigma^2& =& \sigma^2\,,\quad {\rm with}\ \ \sigma >0,\ \beta\geq 0\,.
\label{3rel}
\eea
Note from (\ref{3rel}) that $\Big(A-(s - s_{\!{}_0}) \Sigma\Big)^2= \beta^2$. Also, 
$\displaystyle\frac{d}{d s}\Big(A-(s - s_{\!{}_0}) \Sigma\Big) = \dot A -\Sigma-(s - s_{\!{}_0}) \dot \Sigma = 
A^2 V - (s - s_{\!{}_0}) (\Sigma,A) V = \beta^2\, V$\,.
It is relevant to distinguish the case $\beta=0$ from the rest.
\subsection{The case $\beta=0$}
\label{beta0}
From these last equations we infer that $\beta=0$ implies that $A-(s - s_{\!{}_0}) \Sigma$ is a constant 
lightlike vector, so $A(s_{\!{}_0})$ would be a lightlike vector, which is incompatible with it being orthogonal 
to $V(s_{\!{}_0})$ unless 
$A-(s - s_{\!{}_0}) \Sigma$ vanishes. We conclude that in
this case $A=(s - s_{\!{}_0}) \Sigma$. That is
\bea \dot V &=& (s - s_{\!{}_0}) \Sigma\nonumber\\
\dot \Sigma &=&\sigma^2 (s - s_{\!{}_0}) V
\label{Snap0-2d}
\eea
the second equation coming from the vanishing Snap condition and (\ref{3rel})\,.

We see that this case corresponds to motion in a plane (i.e., 2d) in Minkowski spacetime. 
The solution is
\bea  
V(s) &=&  \cosh(\frac{1}{2}\sigma (s - s_{\!{}_0})^2 )\,V(s_{\!{}_0})+ 
\sinh(\frac{1}{2}\sigma (s - s_{\!{}_0})^2)\frac{1}{\sigma}\Sigma(s_{\!{}_0}) \,,\nonumber \\
 \Sigma(s) &=&  \sigma\sinh(\frac{1}{2} (s - s_{\!{}_0})^2 )\,V(s_{\!{}_0})+ 
\cosh(\frac{1}{2}\sigma (s - s_{\!{}_0})^2)\Sigma(s_{\!{}_0})\label{case-b-0}
\eea
(Note that we may always keep $\sigma>0$) 
If we parametrize the initial conditions 
$$V(s_{\!{}_0}) = \left(
\begin{array}{c}
\cosh(\alpha) \\\sinh(\alpha)\,\vec n_{{}_0}\\
\end{array}
\right)\, \qquad \Sigma(s_{\!{}_0}) =\sigma \left(
\begin{array}{c}
\sinh(\alpha)\\ \cosh(\alpha\,)\,\vec n_{{}_0}\\
\end{array}
\right)
$$
for some real $\alpha$ and a unit norm constant $3$-vector $\vec n_{{}_0}$, 
$\vec n_{{}_0}^2 =1$, then the general solution for $V$ is expressed with an arbitrary second 
degree polynomial with real coefficients $P_2(s) = \frac{1}{2}\sigma (s - s_{\!{}_0})^2 + \alpha$
\beq
V(s) = \left(
\begin{array}{c}
\cosh(P_2(s))\\ \sinh(P_2(s))\,\vec n_{{}_0} 
\end{array}
\right)\, \qquad 
\Sigma(s) = \sigma\left(
\begin{array}{c}
\sinh(P_2(s))\\ \cosh(P_2(s))\,\vec n_{{}_0} 
\end{array}
\right)\,,
\label{2d-sol-Snap-0}
\eeq
with $P_2$ an arbitrary second degree polynomial with real coeffients and with $\vec n_{{}_0}$ a 
normalized constant $3$-vector. In fact we can restrict $\sigma$ to be positive 
by eventually absorbing a minus sign within $\vec n_{{}_0}$ in (\ref{2d-sol-Snap-0}). 
When we send the oberver to rest by using the appropriate boost, we get the constant proper Jerk,
\beq
\Sigma_p(s) =\sigma \left(
\begin{array}{c}
0\\ \vec n_{{}_0} 
\end{array}
\right)\,.
\label{2d-sol-Snap-0-propJerk}
\eeq

Note the change of the degree of the polynomial when compared with (\ref{2d-ct-acc-obs}). 
Now the argument
within the hyperbolic trigonometric functions is a second degree polynomial, instead of the first degree
polynomial which we found for the the solution of the vanishing Jerk case, (\ref{2d-ct-acc-obs}). 
This observation opens a natural way 
to explore 2d solutions for the vanishing Crackle (constant proper Snap) case and higher derivative 
cases in general. This issue will be addressed in section \ref{2d-solved}.
\subsection{The case $\beta\neq 0$}
\label{betano0}
Collecting the expressions in the first part of this section, and defining the vector 
$D = A-(s - s_{\!{}_0}) \Sigma$, we have from the previous analysis the equations
\bea
\dot V &=& A= D+ (s - s_{\!{}_0})\, \Sigma \nonumber \\
\dot D &=& \beta^2 V\nonumber \\
\dot \Sigma &=& \sigma^2 (s - s_{\!{}_0})\, V
\label{VDS}
\eea
with $(V)^2=-1,\ (D)^2 = \beta^2,\ (\Sigma)^2= \sigma^2$ and all three vectors orthogonal. So in this case
the motion takes place in three dimensions.
The equation for the $3$-vector $V$ is obtained from (\ref{VDS}) as
\bea
\dddot V -(\sigma^2 (s - s_{\!{}_0})^2 +\beta^2) \dot V -3\, \sigma^2 (s - s_{\!{}_0}) V =0
\label{Valone}
\eea

\vs

With the proper time
rescaled to $\tau=\sqrt{\sigma}\, (s-s_{\!{}_0}) $, and defining $\displaystyle W= \frac{1}{\beta} D,\ 
Z= \frac{1}{\sigma} \Sigma$,
the system (\ref{VDS}) becomes simplified (with $\displaystyle V'= \frac{d}{d \tau}V$, etc.),
\bea
V'&=& c\, W +\tau Z \nonumber \\
W'&=& c\,  V\nonumber \\
Z'&=& \tau V\,,
\label{VWZ}
\eea
with $\displaystyle c=\frac{\beta}{\sqrt{\sigma}}$ and the orthonormal system $V, W, Z$,
with $(V)^2=-1,\ (W)^2 = 1,\ (Z)^2= 1$.

In this equation, $V, W, Z$ are three orthonormal vectors -$V$ timelike, $W$ and $Z$ spacelike- in an 
$n$-Minkoswki spacetime. In fact, 
since their derivatives in (\ref{VWZ}) are linear combinations of the three vectors themselves, 
we infer that the motion is at most 
$3$-dimensional, and so we continue with a 3d Minkoswki spacetime. A solution for $V(\tau)$ will
be of the type $V(\tau) = V_0(\tau)\,V(0) + V_1(\tau)\,W(0) + V_2(\tau)\,Z(0) $ and since $V(0), W(0), Z(0)$,
span an orthonormal basis, we can take 
$$V(0)=\left(
\begin{array}{c}
1\\0\\0\\
\end{array}
\right)\,, \quad 
W(0)=\left(
\begin{array}{c}
0\\1\\0\\
\end{array}
\right)\,, \quad 
Z(0)=\left(
\begin{array}{c}
0\\0\\1\\
\end{array}
\right)\,.
$$
Analogously to (\ref{Valone}), a single equation for $V$ is obtained from (\ref{VWZ}) as
\beq
 V''' -(\tau^2+c^2)  V' -3 \tau V =0
\label{singleeq}
\eeq
We have not found analytic solutions for (\ref{VWZ}) nor for its consequence (\ref{singleeq}), 
but we can examine nevertheless the behaviour of the solutions in the asymptotic time regime
$|\tau|>>c$ and in 
the ``mid-term'' regime $|\tau|<<c$. 

For large $|\tau|$  (\ref{VWZ}) is approximated by
\bea
V'&=&\tau Z \nonumber \\
Z'&=& \tau V\,,
\label{VWZasympt}
\eea
which is just the system considered in (\ref{Snap0-2d}), that is, the 2d motion (one spatial dimension) 
with constant proper Jerk. In sharp contrast, for $|\tau|<<c$ the system becomes close to 
\bea
V'&=&c\, W \nonumber \\
W'&=& c\, V\,,
\label{VWZsmalls}
\eea
which is nothing but the 2d motion (one spatial dimension) 
with constant proper acceleration (vanishing Jerk), already discussed in (\ref{ctproperA}). 
Therefore (\ref{VWZ}) -and (\ref{VDS})- describes a motion which is composite 
of these two fundamental motions, now in 3d (two spatial dimensions).

  \subsection{Vanishing Snap vs constant proper Jerk}
\label{Jerkbeyond2d}
We have seen in section \ref{EnterstheJerk} that vanishing Jerk implies motion in a plane 
(2d in Minkowski)
and constant proper acceleration. On the other hand, we have seen in the previous section \ref{betano0}
that vanishing Snap does not imply in general 2d motion. 
We may wonder now whether a vanishing Snap condition will at least imply a constant proper Jerk. 
The answer is in the negative. Let us examine it.

To send the observer to the rest frame we will apply the pure boost (\ref{theboost}), equivalently 
written as
\beq B
=\left(
\begin{array}{cccc}
 V_0  & -V_1 \ & -V_2 \ & -V_3 \ \\
 -V_1 \ & 1+\frac{1}{V_0 +1}V_1^2 \ & \frac{1}{V_0 +1}V_1 V_2\ 
 & \frac{1}{V_0 +1}V_1  V_3 \ \\
 -V_2 \ & \frac{1}{V_0 +1}V_1 V_2\ & 1+\frac{1}{V_0 +1}V_2^2\ 
 & \frac{1}{V_0 +1}V_2  V_3 \ \\
 -V_3 \ & \frac{1}{V_0 +1}V_1 V_3\ & \frac{1}{V_0 +1}V_2 V_3\ 
 & 1+\frac{1}{V_0 +1}V_3^2\ \\
\end{array}
\right)\equiv\left(
\begin{array}{cc}
\, V_0 \quad & -\vec V  \\
 -\vec V \quad\ & I+\frac{1}{V_0 +1}\vec V \otimes \vec V   \\
\end{array}
\right)\,.
\label{theboost2}
\eeq

Let us add to the vanishing Snap condition that of the constant proper Jerk. By this we mean that
when we send $\Sigma$ 
to the comoving frame, the resulting object, $\Sigma_p$, is a constant vector,
\beq 
B\,\Sigma = \left(
\begin{array}{c}
0 \\ \vec b_{\!{}_0}
\end{array}
\right)
\label{BSigma}
\eeq
with $\vec b_{\!{}_0}$ a constant $3$-vector. 
We obtain from (\ref{BSigma}) and (\ref{theboost2}), with 
$\Sigma=(\Sigma_0, \vec \Sigma)$,
\beq
\vec \Sigma - \frac{\Sigma_0}{V_0+1} \vec V= \vec b_{\!{}_0} \,,
\label{vecJ}
\eeq
which reads, in $4$-vector language,
$$
\Sigma - \frac{\Sigma_0}{V_0+1} V = \left(
\begin{array}{c}
\frac{\Sigma_0}{V_0+1} \\ \vec b_{\!{}_0}
\end{array}
\right)\,,
$$
which implies, by derivating with respect to the proper time, 
$$
\dot \Sigma - \frac{d}{d s}\Big(\frac{\Sigma_0}{V_0+1}\Big) V - \frac{\Sigma_0}{V_0+1}\dot V = \left(
\begin{array}{c}
\frac{d}{d s}(\frac{\Sigma_0}{V_0+1}) \\ \vec 0
\end{array}
\right)\,,
$$
and using (\ref{VDS}),
\beq
\sigma^2 (s - s_{\!{}_0})\, V - \frac{d}{d s}\Big(\frac{\Sigma_0}{V_0+1}\Big) V
- \frac{\Sigma_0}{V_0+1} \Big(D+ (s - s_{\!{}_0})\, \Sigma\Big)= \left(
\begin{array}{c}
\frac{d}{d s}(\frac{\Sigma_0}{V_0+1}) \\ \vec 0
\end{array}
\right)\,,
\label{usingVDS}
\eeq
The scalar product of this equality with respect to $V, D, \Sigma$, gives, respectively
\beq
-\sigma^2 (s - s_{\!{}_0}) + \frac{d}{d s}\Big(\frac{\Sigma_0}{V_0+1}\Big)
= -\frac{d}{d s}\Big(\frac{\Sigma_0}{V_0+1}\Big) V_0\,,
\label{exp1}
\eeq
\beq
-\frac{\Sigma_0}{V_0+1} \beta^2 = -\frac{d}{d s}\Big(\frac{\Sigma_0}{V_0+1}\Big) D_0\,,
\label{exp2}
\eeq
\beq
-\frac{\Sigma_0}{V_0+1} \sigma^2 (s - s_{\!{}_0}) = -\frac{d}{d s}\Big(\frac{\Sigma_0}{V_0+1}\Big) \Sigma_0\,,
\label{exp3}
\eeq
with 
$D=(D_0, \vec D)$. The content of (\ref{exp1}) and (\ref{exp3}) is the same: 
$\displaystyle \frac{d}{d s}\Big(\frac{\Sigma_0}{V_0+1}\Big) = \frac{\sigma^2 (s - s_{\!{}_0})}{V_0+1}$. 
Injecting this equality 
in (\ref{exp2}) we obtain
\beq
\Sigma_0\,\beta^2= \sigma^2 (s - s_{\!{}_0}) D_0\,.
\label{exp4}
\eeq
Next, taking into account that, from (\ref{VDS}),
$\dot \Sigma_0 = \sigma^2 (s - s_{\!{}_0}) V_0$ and $\dot D_0 = \beta^2 V_0$, we can compute the proper 
time derivative 
of (\ref{exp4}), which yields $D_0=0$. This result implies that 
$\beta =0$ (since $\dot D_0 = \beta^2 V_0$ and $V_0>0$).

We conclude that, within the vanishing Snap case, the requirement of constant proper Jerk 
corresponds to the case $\beta =0$, analyzed in subsection \ref{beta0}, 
which also implies that the motion is 2d. 

\BOXthree{\vfill\vspace{1.4mm}
\it In Minkowski spacetime: 

A trajectory of vanishing Snap has constant proper Jerk if and only if the parameter $\beta$ 
in (\ref{3rel}) vanishes. Then the motion takes place in a plane (2d).
\vfill\vspace{1.5mm}}
\label{eqaccBOX2}

Note however that constant proper Jerk trajectories do not need to be in a plane. As a matter of fact, 
the equation for constant proper Jerk (\ref{vecJ}) can be equivalently written as
\beq
\vec \Sigma - \frac{(\vec b_{\!{}_0} , \vec V)}{V_0+1} \vec V= \vec b_{\!{}_0} \,,
\label{vecJ2}
\eeq
where we have used that $(\Sigma, V)=0$ implies 
$\displaystyle \Sigma_0 =\frac{(\vec \Sigma , \vec V)}{V_0}$ which, 
together with (\ref{vecJ}), gives $\Sigma_0 = \vec b_{\!{}_0} \cdot \vec V$. Also from 
(\ref{vecJ2}) one can go back to (\ref{vecJ}). Since (\ref{vecJ2}) is a second order -non linear- 
differential vector equation in normal form
\beq
\ddot{\vec V} - (A,A)\vec V - \frac{(\vec b_{\!{}_0}, \vec V)}{V_0+1} \vec V- \vec b_{\!{}_0} =0\,,
\label{vecJ3}
\eeq
with $\displaystyle(A,A)= -A_0^2 + (\vec A)^2 = -\Big(\frac{(\dot{\vec V},\vec V)}{V_0}\Big)^2 
+ (\dot{\vec V})^2$, we infer that it has solutions for arbitrary initial conditions $\vec V(t_0)$ and 
$\dot{\vec V}(t_0)$.
Thus the solution trajectories of (\ref{vecJ3}) are not restricted to be in a plane in Minkowski spacetime.
\vs
\subsection{Some equivalences.}
\label{someq}
Our findings can be summed up with the following equivalences. Let us introduce the three sentences:

{\bf A:} The trajectory has vanishing Snap,

{\bf B:} The trajectory has constant proper Jerk,

{\bf C:} The trajectory takes place in a plane (2d) in Minkowski spacetime,

\vs

then what we have shown in subsection \ref{Jerkbeyond2d} is that 

{\bf 1)}\ {\bf A}\,$\cap$\,{\bf B} $\Rightarrow$ {\bf C}\,.

We can also show that  

{\bf 2)}\ {\bf B}\,$\cap$\,{\bf C} $\Rightarrow$ {\bf A}\,, 

and that 

{\bf 3)}\ {\bf A}\,$\cap$\,{\bf C} $\Rightarrow$ {\bf B}\,.

\vspace{5mm}

The proofs of {\bf 2)} and {\bf 3)} are postponed to section \ref{2d-solved} 
in which we show that in 2d, the 
notions of vanishing Snap and constant proper Jerk coincide. See in particular the equations 
(\ref{2d-V-generalSI}) and (\ref{2d-Vn-proper}) which prove the more general statement that in 2d 
the notion of vanishing $(n+1)$-acceleration and constant proper $n$-acceleration coincide.

 Summarizing:
\BOXfour{\vfill\vspace{1.4mm}
\it In Minkowski spacetime: 

\hspace{1.6cm} {\bf A}\,$\cap$\,{\bf B}\,=\,{\bf B}\,$\cap$\,{\bf C}\,=\,{\bf A}\,$\cap$\,{\bf C}\,=
{\bf A}\,$\cap$\,{\bf B}\,$\cap$\,{\bf C}\,.
\vfill\vspace{1.5mm}}
\label{eqaccBOX3}


 
\section{The view from another frame of reference}
\label{CpAJS} 
Consider a motion with constant proper acceleration, that is, satisfying (\ref{propA}) (with constant 
$\vec a$) or, equivalently, (\ref{eqpropaccproptime}) {\sl in a given frame of reference}. 
If our Minkowski spacetime has dimension $>2$, we may wonder how this motion is seen from a
different frame of reference, 
and in particular what will be the proper acceleration in the instantaneous rest frame of the observer.

So consider a general 2d motion in a 4d (coordinates $t,x,y,z$)
Minkowski spacetime -although 3d is enough to see what happens.  
Choose coordinates so that this rectilinear motion takes place in the $t,x$ coordinates,
$ 
V =(V_0, \vec V) = (V_0(s), V_1(s),0,0) $,
and let us apply to it a proper boost $G$ characterized by sending a given timelike 
unit-norm constant vector
$ U =(U_0, \vec U )=(U_0, U_1 , U_2,0 )$  to rest, $(1,0,0,0)$, that is,
$$
G= \left(
\begin{array}{cc}
\, U_0 \quad & -\vec U  \\
 -\vec U \quad\ & I+\frac{1}{U_0 +1}\vec U \otimes \vec U   \\
\end{array}
\right)\,.
$$
After this change of reference frame, the new velocity of the observer will be $W(s) =: G\, V(s)$ and 
there will be an instantaneous proper boost $B_W(s)$\footnote{$B_W(s)$ is just the proper 
boost (\ref{theboost2}) with $V$ replaced by $W$.} which sends this velocity to the instantaneous
rest frame, $B_W(s)\, W(s)=B_W(s)\, G\, V(s)= (1,0,0,0) = V_p$. 
But we know that $B(s)\, V(s)= (1,0,0,0) = V_p$. So $B_W(s)\, G\, B(s)^{-1}$ must be a spatial rotation 
$R(\theta(s))$,
\beq
B_W(s)\, G\, B(s)^{-1} \equiv R(s)= \left(
\begin{array}{cc}
\, 1 \quad & 0  \\
 0 \ \ & R_{ij}(\theta(s))   \\
\end{array}
\right)\,.
\label{R(s)}
\eeq
This rotation acts on the $2$-dimensional plane $<x,y>$ and has $\cos(\theta)$ determined as 
\beq\cos(\theta(s)) = 1-\frac{{(U_2)}^2 ({V_0}-1)}{({U_0}+1) (1-(U \cdot V))}\,,
\label{theRot}
\eeq
with $U \cdot V=  - U_0\,V_0+\vec U \cdot\vec V$. The axis of 
rotation of (\ref{R(s)}) is in the direction $\vec U \wedge\vec V$. 

Note that Eq.(\ref{theRot}) is symmetric with respect to $U$ and $V$. Indeed, since  
$$\displaystyle  (U_2)^2 = \frac{|\vec U \wedge\vec V|^2 }{|\vec V|^2} = |\vec U|^2 (\sin\alpha)^2= 
(U_0^2 -1) (\sin\alpha)^2\,,
$$
where $\alpha$ is the angle between $\vec V$ and $\vec U$, then, 
in terms of $\displaystyle \frac{\theta}{2}$ equation (\ref{theRot}) becomes 
\beq
(\sin\frac{\theta}{2})^2 = \Big(\frac{(U_0 -1)(V_0 -1)}{2\,(1-(U \cdot V))}\Big)\,(\sin\alpha)^2\,,
\label{theAng}
\eeq
(note that there is an $\alpha$ dependence in $(U \cdot V)$ as well) 
and thus the rotation $R_{ij}(\theta)$ becomes the identity only for $\sin\alpha=0$, 
which is equivalent to the vanishing of $U_2$ -when $\vec U$ and $\vec V$ are parallel.

When the rotation is not the identity it can still be a constant rotation (that is, $s$-independent)
only if $V_1$ is constant, that is, when the rectilinear motion of the observer is uniform. But not
for the accelerated observer.

Note on the other hand that the proper acceleration, as seen from the new reference frame, is
($[p]$ is for ``proper''),
$A_{W}^{[p]}= B_W(s)\, A_W = B_W(s)\, G\, A = B_W(s)\, G\, B(s)^{-1} A^{[p]} = R(\theta(s)) A^{[p]}$\,. 
So for a constant proper acceleration $A^{[p]}=(0,a_{\!{}_0},0,0)$, we end up with 

\beq
A_p(s) = \left(
\begin{array}{c}
0\\a_{\!{}_0}\\0\\0\\
\end{array}
\right)\  \longrightarrow\ A_{W}^{[p]} =a_{\!{}_0}\left(
\begin{array}{c}0\\
\cos(\theta(s))\\ \sin(\theta(s))\\0\\
\end{array}
\right) \,.
\label{}
\eeq

Thus, if we are not in the case of linear uniform motion for the observer, only when the component 
$U_2 $ vanishes will this 
rotation be independent of the component $V_1(s)$, in which case the rotation becomes 
the identity matrix. The boost $G$ in this case will be along the $x$ direction, thus commuting
with the family of boosts $B(s)$, which are already in such direction.

This argument applies as well to any motion (not necessarily 2d) 
with constant proper Jerk, or constant proper Snap, etc. 
The eventual constancy of these objects is frame dependent and only for changes of frame engendered by 
the boost $G$ when it points in 
the direction of the $3$-velocity of the observer (boost in the $2$-dimensional plane $<t,x>$ 
  in our case), will this constancy be preserved. Indeed, we should rephrase the notion of 
  constant proper Jerk as: 
{\sl a trajectory has constant proper Jerk in Minkowski spacetime if there exists a frame of 
reference such that equation (\ref{vecJ}) holds for some constant vector $\vec b_{\!{}_0}$\,. 
}

\vs

The conclusion is clear and general: the notion of constant proper acceleration, constant proper 
Jerk, constant proper Snap, etc, 
as defined in (\ref{BSigma}) for the Jerk, and trivially generalized to all these objects, 
is not invariant under changes of reference frame. On the other hand, it is reasonable to consider 
that the ``constancy'' of some $A_n$ should be associated with the vanishing of the next $A_{n+1}$. 
For that we need the additional requirement 
(see section \ref{someq} for the case of the constant proper Jerk)
that the trajectory takes place in a 2d plane in Minkowski spacetime.

All in all, and in view to eventual generalizations to curved spacetime, our statements
should at least be invariant under changes of frame of reference, with the expectation for them 
to generalize to statements invariant under diffeomorphisms in generally covariant theories. 
If we want to make invariant statements we should rely in the curvature invariants 
associated with the trajectory. These curvature invariants  will be introduced in the context of  
the Frenet-Serret formalism. This formalism and its relation with the objects introduced 
so far (Jerk, Snap, etc.), is discussed in the next section.

\section{The Frenet-Serret framework}
\label{FS}

The Lorentzian setting for the originally Euclidean Frenet-Serret formalism was studied 
in \cite{Letaw:1980yv}.
Here we directly adapt to the Lorentzian setting the Euclidean construction in \cite{Ramos:1995zr}. 
Consider a trajectory $X(s)$ parametrized by the proper time $ s$, so that 
the velocity vector $U_0$, defined as $\displaystyle U_0\equiv V=\frac{d}{d\,s} X $, has $U_0^2 = -1$.

The first curvature invariant (also known as scalar curvature) is defined by 
$\displaystyle\kappa_1=|\frac{d}{d\,s}U_0|$, and the vector
$U_1$, orthogonal to $U_0$ -and hence spacelike- and normalized, is defined as,
$$
U_1 = \frac{1}{\kappa_1}\,\frac{d}{d\,s} U_0\,,\qquad U_1^2 =1\,.
$$
Of course this construction only works for non-vanishing $\kappa_1$. There may be exceptions at 
some isolated points in the trajectory, but if $\kappa_1=0$ for an entire segment of the trajectory, 
then the construction finishes here -for this segment. Note that the acceleration $A$ defined in 
previous sections (see (\ref{asexpected}),(\ref{theA}))
is just $\displaystyle A= \kappa_1\, U_1$ (so $|A|= \kappa_1$). 

Next, noticing that the vector $\displaystyle\frac{d}{d\,s} U_1 -\kappa_1\,U_0$ is orthogonal to both
$U_0$ and $U_1$, we define the second curvature invariant (also known as the torsion scalar)
$$
\kappa_2 =\Big|\frac{d}{d\,s} U_1 -\kappa_1\,U_0\Big|
$$
and as long as $\kappa_2>0$ (otherwise the construction finishes here) we define the spacelike 
normalized vector $U_2$ as
$$
U_2 = \frac{1}{\kappa_2}\,\Big(\frac{d}{d\,s} U_1 -\kappa_1\,U_0\Big)\,,
$$
which is an expression that can be read also as
$$
\frac{d}{d\,s} U_1 = \kappa_2\,U_2 + \kappa_1\,U_0,\qquad  U_2^2 =1,\ (U_1\cdot U_2)=0,
\ (U_0\cdot U_2)=0\,.
$$
Again, having noticed that the vector $\displaystyle\frac{d}{d\,s} U_2 +\kappa_2\,U_1$ is orthogonal to 
$U_0$, $U_1$ and $U_2$, we define the third curvature invariant (also known as the hypertorsion scalar)
$$
\kappa_3 =\Big|\frac{d}{d\,s} U_2 +\kappa_2\,U_1\Big|\,,
$$
alongside with the definition of $U_3$ (as long as $\kappa_3>0$, otherwise the construction finishes here)
$$
U_3=  \frac{1}{\kappa_3}\Big(\frac{d}{d\,s} U_2 +\kappa_2\,U_1\Big)
$$
which is an expression that can be read also as
$$
\frac{d}{d\,s} U_2 = \kappa_3\,U_3 -\kappa_2\,U_1,\qquad U_3^2 =1,\ (U_2\cdot U_3)=0,\ (U_1\cdot U_3)=0,
\ (U_0\cdot U_3)=0\,.
$$
and this construction continues as long as it is allowed by the number of spacetime of dimensions and 
the specific curvatures of the trajectory considered. In 4d Minkowski we have reached the full constuction.

Note that these curvature invariants, although constructed for trajectories in Minkowski spacetime, 
can be generalized to curved spacetimes along the lines of section \ref{Extension}. 
In this broader sense, these curvature invariants 
behave as scalars under target space diffeomorphisms (coordinate
reparametrizations) and as invariants under world line diffeomorphisms (reparametrizations of the 
evolutionary parameter of the trajectory). The derivatives of these curvature invariants with respect
to the proper time are also curvature invariants with the same properties with respect 
to both types of diffeomorphisms as the original curvature invariants discussed above.

\subsection{Revisiting the constantly accelerated observer}
\label{revisit}
By the reasons  given in section (\ref{CpAJS}). the solution given in section \ref{the3dsol} 
for the 3d constantly accelerated observer does not show constant
proper acceleration when examined form another generic frame of reference. 
As said before, this is not a real problem for the claim that this is a solution for the trajectory
of a constantly accelerated observer, but we can take advantage of the Frenet-Serret
formalism of section (\ref{FS}) to show that it is possible to characterize this solution in an 
invariant way. 

One can easily compute the first and second curvature scalars for the solution (\ref{V-3d}). By
construction we clearly have $\kappa_1 = a_{\!{}_0}$. The second curvature scalar turns out to be 
$\kappa_2= a_{\!{}_0}\,\sqrt{1-d}$\, (Thus $\kappa_2<\kappa_1$). 
Since we are in 3d Minkowski spacetime, $\kappa_3$ vanishes.
With these data we can reconstruct the solution (\ref{V-3d}), as we do now.

\vs

For constant values of $\kappa_1\,,\kappa_2\,,\kappa_3$ \,, 
the solution trajectories were classified in \cite{Letaw:1980yv}.
Using the Frenet Serret equations of section (\ref{FS}), being $\kappa_1$ and $\kappa_2$ constants
(and $\kappa_3 =0$), we obtain the equation ($U_0=V$)
\beq\dddot V = (\kappa_1^2-\kappa_2^2)\, \dot V\,,
\label{k1k2const}
\eeq
and in our case $\kappa_1^2-\kappa_2^2=a_{\!{}_0}^2\,  d >0$. The fact that 
$\kappa_2<\kappa_1$ is crucial.
The general solution for  $V$ is then of the form
$$
V = M \cosh(\sqrt{d}\,a_{\!{}_0}(s-s_{\!{}_0})) + N  \sinh(\sqrt{d}\,a_{\!{}_0}(s-s_{\!{}_0})) + G\,,
$$
with $M,\,N,\,G$ constant vectors to be determined.
The condition $(V)^2=-1$ imposes $(M,N) = (M,G) = (N,G) = 0, \ (N)^2 =-(M)^2$ and $(M)^2+(G)^2 =-1$.
The acceleration is 
$$
A=\dot V = \sqrt{d}\,a_{\!{}_0} \Big(M \sinh(\sqrt{d}\,a_{\!{}_0}(s-s_{\!{}_0})) + 
N  \cosh(\sqrt{d}\,a_{\!{}_0}(s-s_{\!{}_0}))\Big),
$$
and $(A)^2= a_{\!{}_0}^2$ imposes $\displaystyle (M)^2 = -\frac{1}{d}$ which implies 
$\displaystyle (N)^2 = \frac{1}{d}$ and $\displaystyle (G)^2 = \frac{1-d}{d}$\,.

With all these restrictions we can choose $M,N,G$. For instance the choice made in (\ref{V-3d}) is
$$
M=\left(
\begin{array}{c}
\frac{1}{d}\\ 0 \\ \frac{\sqrt{1-d}}{d}
\\
\end{array}
\right)\,,\qquad N=\left(
\begin{array}{c}
0 \\ \frac{1}{\sqrt d} \\ 0
\\
\end{array}
\right)\,,\qquad G=\left(
\begin{array}{c}
\frac{1-d}{d} \\ 0 \\ \frac{\sqrt{1-d}}{d}
\\
\end{array}
\right)\,.
$$
Another, simpler choice, related to the one above by a boost, is  
$$
M=\left(
\begin{array}{c}
\frac{1}{\sqrt d}\\ 0 \\ 0
\\
\end{array}
\right)\,,\qquad N=\left(
\begin{array}{c}
0 \\ \frac{1}{\sqrt d} \\ 0
\\
\end{array}
\right)\,,\qquad G=\left(
\begin{array}{c}
0 \\ 0 \\ \sqrt{\frac{1-d}{d}}
\\
\end{array}
\right)\,,
$$
with which the solution (\ref{V-3d}) takes the form
\beq
V = \left(
\begin{array}{c}
V_0 \\ \vec V \\
\end{array}
\right) = \left(
\begin{array}{c}
\frac{1}{\sqrt d} \cosh(\sqrt{d}\,a_{\!{}_0} (s-s_{\!{}_0})) 
\\ \frac{1}{\sqrt d}\sinh(\sqrt{d}\,  a_{\!{}_0} (s-s_{\!{}_0})) \\  
\sqrt{\frac{1-d}{d}}\\
\end{array}
\right)\,.
\label{V-3dbis}
\eeq
Interestingly we may interpret it as the standard accelerated motion along the $x$ direction 
together with a motion along the $y$ direction which is 
{\sl linear in the proper time}. Note however that the acceleration $A$ associated 
with this solution (\ref{V-3dbis}) does not give a constant proper acceleration, whereas the solution
(\ref{V-3d}) does, because of the frame dependence explained in section \ref{CpAJS}. But we can 
nevertheless still claim that (\ref{V-3dbis}) describes a motion with constant proper acceleration.

Since the coordinate time satisfies
$\displaystyle \frac{d\,t}{d\,s} = V_0 = 
\frac{1}{\sqrt d} \cosh(\sqrt{d}\,a_{\!{}_0} (s-s_{\!{}_0}))$, we can set
$\displaystyle t= \frac{1}{a_{\!{}_0}d} \sinh(\sqrt{d}\,a_{\!{}_0} \,(s-s_{\!{}_0}))$ 
(so $s=s_{\!{}_0}$ is $t=0$) and the position vector becomes, in terms of the coordinate time,
\beq
X =X(0)+  \left(
\begin{array}{c}
t \\ \frac{1}{a_{\!{}_0}\,d}(\sqrt{1+ a_{\!{}_0}^2\,d^2\, t^2}-1)\\ 
\frac{\sqrt{1-d}}{a_{\!{}_0}\,d}\,{\rm arcsinh}(d\, a_{\!{}_0}\,t)   \\
\end{array}
\right)
\label{X-3d}
\eeq

We notice that this reconstruction relies on the fact that, in addition to the constancy of 
$ \kappa_1,\, \kappa_2$, the inequality $\kappa_2<\kappa_1$ is satisfied. We may state

\BOXthree{\vfill\vspace{1.4mm}
\it In Minkowski spacetime (any dimension), the observer with 
constant proper acceleration is characterized 
by $\kappa_3=0$ and $ \kappa_1,\, \kappa_2$ constants with $\kappa_2<\kappa_1$. 
The particular case $\kappa_2=0$ corresponds to motion in a plane and it also identifies 
the case of the vanishing of the Jerk.
}
\label{cpa-car}
These  statements can be exported as definitions in curved spacetime, 
just by dropping the reference to the trajectory taking place in a plane,
\BOXthree{\vfill\vspace{1.4mm}
\it In curved spacetime, the observer with 
constant proper acceleration is defined  
by $\kappa_3=0$ and $ \kappa_1,\, \kappa_2$ constants with $\kappa_2<\kappa_1$.  
The particular case $\kappa_2=0$ corresponds to the vanishing of the Jerk.}
\label{cpa-car-curv}

\vs

Although both solutions, (\ref{V-3d}) and (\ref{V-3dbis}), are related by a simple boost, 
it is worth to notice that the 
trajectory (\ref{X-3d}), unlike (\ref{XV-3d}), does not have asymptotes in the (x,y) plane.
Now the ingoing and  outgoing directions in the $2$-space are just oposite: $(-1,0),\ \ (1,0)$. 
The parameter $d$ is
now associated with the ratio of the two space coordinates at infinity. Indeed we have,
for (\ref{X-3d}), 
($X=\{t,x,y\}$),  
$$\lim_{s\to\infty}\frac{y}{\log x}= \frac{\sqrt{1-d}}{a_{\!{}_0}^2\,d}\,.
$$

\vs


\vs

For the sake of completeness we may consider the cases with $\kappa_2\geq \kappa_1$. 
In view of \ref{k1k2const}, the case $\kappa_2= \kappa_1$ is immediate. 
The case with $\kappa_2>\kappa_1$ corresponds to uniform circular motion, 
which we review briefly in the Appendix \ref{A} and give expressions for its $n$-accelerations.

\subsection{ Jerk and Snap in terms of the Frenet-Serret basis}
 \label{JS-FS}
In terms of the Frenet-Serret basis, we can write the Jerk and Snap as
\beq
\Sigma = \frac{d \kappa_1}{d\,s} U_1 + \kappa_1 \kappa_2 U_2\,,
\label{jerk-FS}
\eeq
\beq
\Xi = \Big(\frac{d^2 \kappa_1}{d\,s^2}-\kappa_1 \kappa_2^2\Big) U_1 
+ \Big(\frac{d }{d\,s}(\kappa_1^2\kappa_2)\Big)U_2
+\Big(\kappa_1\kappa_2\kappa_3\Big)U_3\,.
\label{snap-FS}
\eeq
Now we are ready to reproduce the results of sections \ref{EnterstheJerk} and \ref{vanishingSnap}. 

\vspace{3mm}

Assuming that the acceleration is non vanishing, $\kappa_1> 0$ (otherwise the Jerk already
vanishes), vanishing of the Jerk ($\Sigma=0$),
means $\kappa_2=0$ and $\displaystyle\frac{d \kappa_1}{d\,s}=0$. 
Here we recover the results of sction \ref{EnterstheJerk}. 
Whereas $\kappa_2=0$ signals that the 
trajectories lie in a plane, the second 
condition, integrated to  $\kappa_1=|A|=$ constant, is the well know equation (\ref{eqacc}) 
for the constantly accelerated observer (hyperbolic trajectories) in Minkowski 2d.

\vspace{3mm}

Vanishing of the Snap is equivalent to 
$$
\Xi=0\quad  \Leftrightarrow\quad  \kappa_3=0,\quad \kappa_1^2\kappa_2={\rm constant},
\quad \frac{d^2 \kappa_1}{d\,s^2}-\kappa_1 \kappa_2^2 =0.
$$
Solving these equations we obtain for $\kappa_1\,,\,\kappa_2$, 
in terms of the notation of section \ref{vanishingSnap}, 

\beq
\kappa_1= \sqrt{\sigma^2(s-s_{\!{}_0})^2 + \beta^2}\,,\qquad 
\kappa_2=\frac{\beta\,\sigma}{\sigma^2(s-s_{\!{}_0})^2 + \beta^2}\,.
\label{k1k2}
\eeq
with $\sigma>0\,,\, \beta\geq 0$ (for $\sigma=0$ we are back to the vanishing Jerk case). 
In terms of the orthogonal vectors $V,\, D,\, \Sigma$, 
used in section  
\ref{vanishingSnap}, we have ($U_2$ only exists if $\beta>0$),
\beq U_0=V, \qquad 
U_1 =\frac{1}{\kappa_1}\Big(D + (s-s_{\!{}_0}) \Sigma\Big),\qquad
U_2 = \frac{1}{\beta\,\sigma\,\kappa_1}\Big(-\sigma^2(s-s_{\!{}_0})D + \beta^2 \Sigma\Big)\,.
\label{basisch}
\eeq

Notice from (\ref{k1k2}) that
$$
\kappa_2= 0 \quad  \Leftrightarrow\quad \beta=0\,.
$$
This is the invariant statement we were after: we showed in section \ref{vanishingSnap}
that within
the vanishing Snap condition, constant proper Jerk was equivalent to $\beta=0$. On the other hand, 
the constancy of the proper Jerk does not imply the vanishing of the Snap.

We are led to the following statement

\vs

\BOX{\vfill\vspace{1.4mm}
\it In Minkowski spacetime: 

A trajectory of vanishing Snap has constant proper Jerk only if $\kappa_2= 0$ 
(thus implying $\kappa_i= 0,\,i>2$) and $\frac{d^2 \kappa_1}{d\,s^2}=0$. 
This trajectory is therefore 2d and $\kappa_1$ is (the absolute value of) 
an arbitrary first degree polynomial in the proper time.
\vfill\vspace{1.5mm}}
\label{eqaccBOX4}

This satement can be taken as a definition in curved spacetime

\vs

\BOX{\vfill\vspace{1.4mm}
\it In curved spacetime: 

A trajectory of vanishing Snap and constant proper Jerk is characterized by $\kappa_2= 0$ 
and $\frac{d^2 \kappa_1}{d\,s^2}=0$. 
\vfill\vspace{1.5mm}}
\label{eqaccBOX4curv}

Let us take stock of our results so far. We have seen that for $n=1$ and $n=2$, the concepts of 
constant $n$-acceleration and vanishing $(n+1)$-acceleration only coincide if the motion is 2d 
in Minkowski spacetime or, more generally, either in flat or curved spacetime, if $\kappa_2=0$. 
On the other hand we will see in section \ref{2d-solved} that if $\kappa_2$ vanishes then both 
concepts coincide $\forall n$. It is plausible that, as in the  $n=1$ and $n=2$ cases, this coincidence 
only happens when $\kappa_2=0$. 

In the case of vanishing Snap, the fact that the curvature invariants are no longer constants is
the circumstance that makes it difficult to find analytic solutions. In addition, we do not have 
analytic solutions for the constant proper Jerk case nor for its characterization through
curvature invariants.

Up to now, we have shown that the first, second and third accelerations $A_1=A\,,\ A_2=\Sigma\,,\ A_3=\Xi$, 
have a common structure, displayed below. It is not difficult to see that this common structure is 
shared by the whole family of $n$-accelerations, $A_n$,
$$
\displaystyle A_{n+1}= \kappa_1^{(n)}\,U_1 +\ terms\ vanishing\ for\  
\kappa_2=0\,,
$$
where $\displaystyle \kappa_1^{(n)}=\frac{d^{n}}{d\,s^{n}}\kappa_1$. 

\vs

Considering this structure, and anticipating the results of section \ref{2d-solved}, we may state that

\BOX{\vfill\vspace{1.4mm}
\it In flat or curved spacetime: 

If the second curvature invariant vanishes, $\kappa_2 =0$, then the concepts of vanishing 
$(n+1)$-acceleration and constant proper $n$-acceleration are equivalent and are characterized by
$\kappa_1^{(n)}=0$. This trajectory is therefore 2d and $\kappa_1$ is (the absolute value of) an arbitrary
$(n-1)$-degree polynomial in the proper time.

\vfill\vspace{1.5mm}}.
\label{eqaccgen}

\vs

%
%

%
%
  
  \section{2d Minkowski: Constant proper Jerk, Snap, Crackle...}
 \label{2d-solved}
 
In this section we will obtain the solutions for 
the constant proper $n$-acceleration under the assumption that the $(n+1)$-acceleration vanishes. 
Since we have shown in the previous section that in this case the trajectories are $2$-dimensional, 
we will work in a $2$-dimensional Minkowski spacetime.
 
In a 2d Minkowski spacetime we can easily generalize the results in 
(\ref{2d-sol-Snap-0}). As stated in section \ref{JerkSnapCrackle}, we use the notation $V$ 
for the velocity 
(normalized to $(V)^2=-1$) 
, $A_1$
for the acceleration $A$; $A_2$ for the $2$-acceleration or Jerk, $\Sigma$; 
$A_3$ for the $3$-acceleration or Snap, $\Xi$; etc.

With the general parametrization 
\beq
V(s) = \left(
\begin{array}{c}
\cosh(f(s))\\ \sinh(f(s)) \nonumber\\
\end{array}
\right)
\label{2d-V-general}\,,
\eeq
we obtain
\beq
 A_1(s) = f'(s) \left(
\begin{array}{c}
\sinh(f(s))\\ \cosh(f(s)) \nonumber\\
\end{array}
\right)
\,,\qquad  
A_2(s) = f''(s) \left(
\begin{array}{c}
\sinh(f(s))\\ \cosh(f(s)) \nonumber\\
\end{array}
\right)
\label{2d-V-generalbis}\,,
\eeq
(note that $\kappa_1= |f'(s)|$) and in general,
\beq
 A_n(s) = f^{(n)}(s) \left(
\begin{array}{c}
\sinh(f(s))\\ \cosh(f(s))
\end{array}
\right)
\label{2d-V-generalSI}\,,
\eeq
that is, $|\kappa_1^{(n-1)}(s)|= |f^{(n)}(s)|$.  
All this information can be examined at any proper time $s$ from the instantaneous rest system of the 
observer, through the boost
\beq B(s)=\left(
\begin{array}{cc}
\  \cosh(f(s))\  & -\sinh(f(s)) \\
 -\sinh(f(s))\ &\,\ \cosh(f(s)) \\ 
\end{array}
\right)\,,
\label{2d-boost}\,.
\eeq
which sends $V$ to the rest frame -at this proper time $s$- and sends 
$A_1$ to the proper acceleration,
$A_2$ to the proper Jerk and in general sends $A_n$ to its proper expression 
$A_n^{[p]}$
($[p]$ is for ``proper'')
in the rest frame, $A_n(s) \to B(s)A_n(s) = A_n^{[p]}(s)$. 
Thus we have
\beq
V^{[p]}(s) =  \left(
\begin{array}{c}
1\\ 0
\end{array}
\right),\quad 
 A_n^{[p]}(s) = f^{(n)}(s) \left(
\begin{array}{c}
0\\ 1
\end{array}
\right)\,{\rm for} \ n>1
\label{2d-Vn-proper}\,.
\eeq
We conclude that in 2d Minkowski, a constant proper $A_n^{[p]}$ corresponds to $f$ being a 
$n$-degree polynomial $f(s) = P_{n}(s)$ and it also corresponds to a vanishing 
$A_{(n+1)}$. 

\BOXfour{\vfill\vspace{1.4mm}
\it  In 2d Minkowski, constant proper $A_n^{[p]}$\ $\Leftrightarrow$ 
vanishing 
$A_{(n+1)}$\ $\Leftrightarrow \ f(s) = P_{n}(s)\ \Leftrightarrow$\  \ $\kappa_1= |P'_{n}(s)|$
\vfill\vspace{1.5mm}}
\label{2d-general-res}

The fact that $f(s)$ is a $n$-degree polynomial in proper time was already obtained in 
\cite{Russo:2009yd}. Now we provide analytic solutions for the 
trajectories in the simplest case $\displaystyle P_n(s) = \frac{s^n}{s_{\!{}_0}^n}$. 
This case captures the asymptotic behaviour at $s\to \pm \infty$. The trajectories are
\beq
 X(s) =  \left(
\begin{array}{c}
X_0(s)\\ X_1(s)
\end{array}
\right)
\label{2d-V-generalAN}\,.
\eeq
with 
$$
X_0(s)= s \, _1\!F_2\left(\frac{1}{2 n};\frac{1}{2},1+\frac{1}{2 n};\frac{1}{4} 
(\frac{s}{s_{\!{}_0}})^{2 n}
\right)\,,
$$
$$X_1(s)=\frac{1}{n+1} s\,(\frac{s}{s_{\!{}_0}})^{n} \, _1\!F_2\left(\frac{1}{2}+\frac{1}{2
   n};\frac{3}{2},\frac{3}{2}+\frac{1}{2 n};\frac{1}{4}(\frac{s}{s_{\!{}_0}})^{2 n}\right)\,,
$$
where $_1\!F_2$ is the generalized hypergeometric function,
$$
_1\!F_2\left(a;b,c;z\right) = \sum_{k=0}^{\infty} \frac{(a)_{\!{}_k}}{(b)_{\!{}_k} (c)_{\!{}_k}} 
\frac{z^k}{k!}, \quad (x)_{\!{}_k}:= \frac{\Gamma(x+k)}{\Gamma(x)}\,,
$$
with $\Gamma$ the Gamma function. 

To depict the trajectories as tangent to the future light cone with vertex at the origin of coordinates
we must make a translation: 
\beq 
X_1(s)\to X_1(s) + \Gamma(\frac{n+1}{n}) s_{\!{}_0} =: \bar X_1(s)\,. 
\label{transl}
\eeq
This quantity, 
$\displaystyle\Gamma(\frac{n+1}{n}) s_{\!{}_0}$,\ is the equal-time distance between the observer
at $t=0$ and the origin of coordinates. This origin of coordinates belongs to the light cone line 
defining the horizon for the observer. 

Still keeping with $\displaystyle P_n(s) = \frac{s^n}{s_{\!{}_0}^n}$, if we compare the -infinite- time of travel, 
from past infinity 
$s\to -\infty$ to future infinity $s\to \infty$, of the trajectory corresponding to 
$\displaystyle P_n(s)$ and that of the light (considering in the
$n =$ odd case that the light is reflected at the origin of coordinates), we find a finite difference
$d_n$. 
This time delay is 
$$d_n=\displaystyle 2\,\Gamma(\frac{n+1}{n}) s_{\!{}_0}\,.$$ 
It is maximum for $n=1, d_1=2\, s_{\!{}_0}$, that is, for the 
constantly accelerated observer, whereas the minimum is reached for $n=2$,  
$\displaystyle d_2=2\,\Gamma(\frac{3}{2})\, s_{\!{}_0}$. After this minimum, $d_n$ increases 
monotonically with $n$, reaching $d_1$ in the limit $n\to \infty$. 

\vspace {4mm}

\begin{figure}\begin{center}
\includegraphics[width=6cm,angle=0]{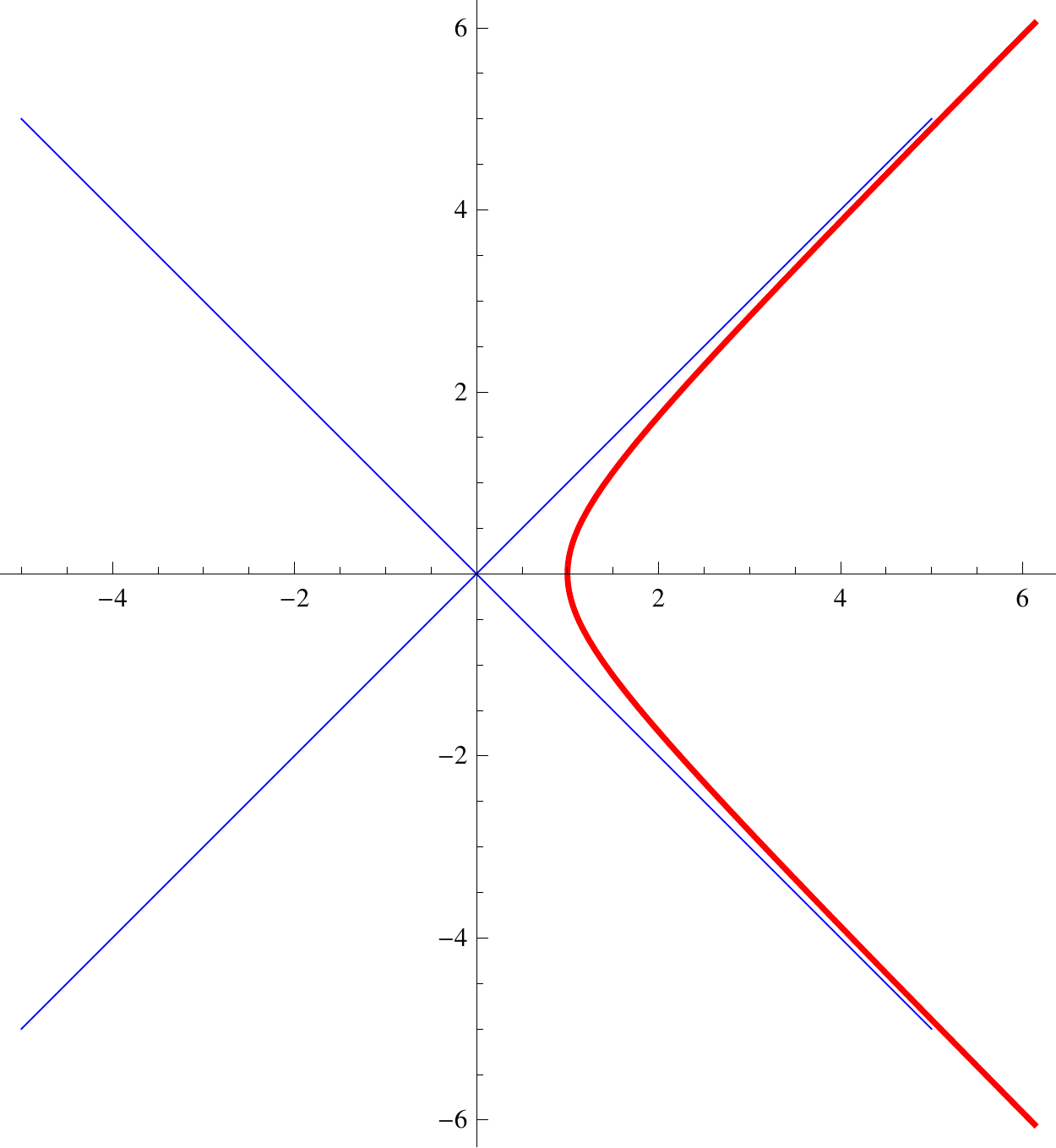}
\end{center}\caption{\sl 2d Minkowski. $n=1$, $\displaystyle P_1(s) = s$. 
The well known hyperbola depicting the trajectory of the observer with vanishing Jerk $=$ 
constant proper acceleration. 
The line $x=t$ of the light cone is a horizon for the observer. This horizon appears in all 
the cases $n\geq 1$. Notice that, as it will happen
with all $n=$ odd cases, the trajectory in the space coordinate $x$ goes from $+\infty$ in the past
to $+\infty$ in the future, reaching the minimum at $t=0$ (in our parametrization).} 
\label{F1}
\end{figure}

\begin{figure}\begin{center}
\includegraphics[width=6cm,angle=0]{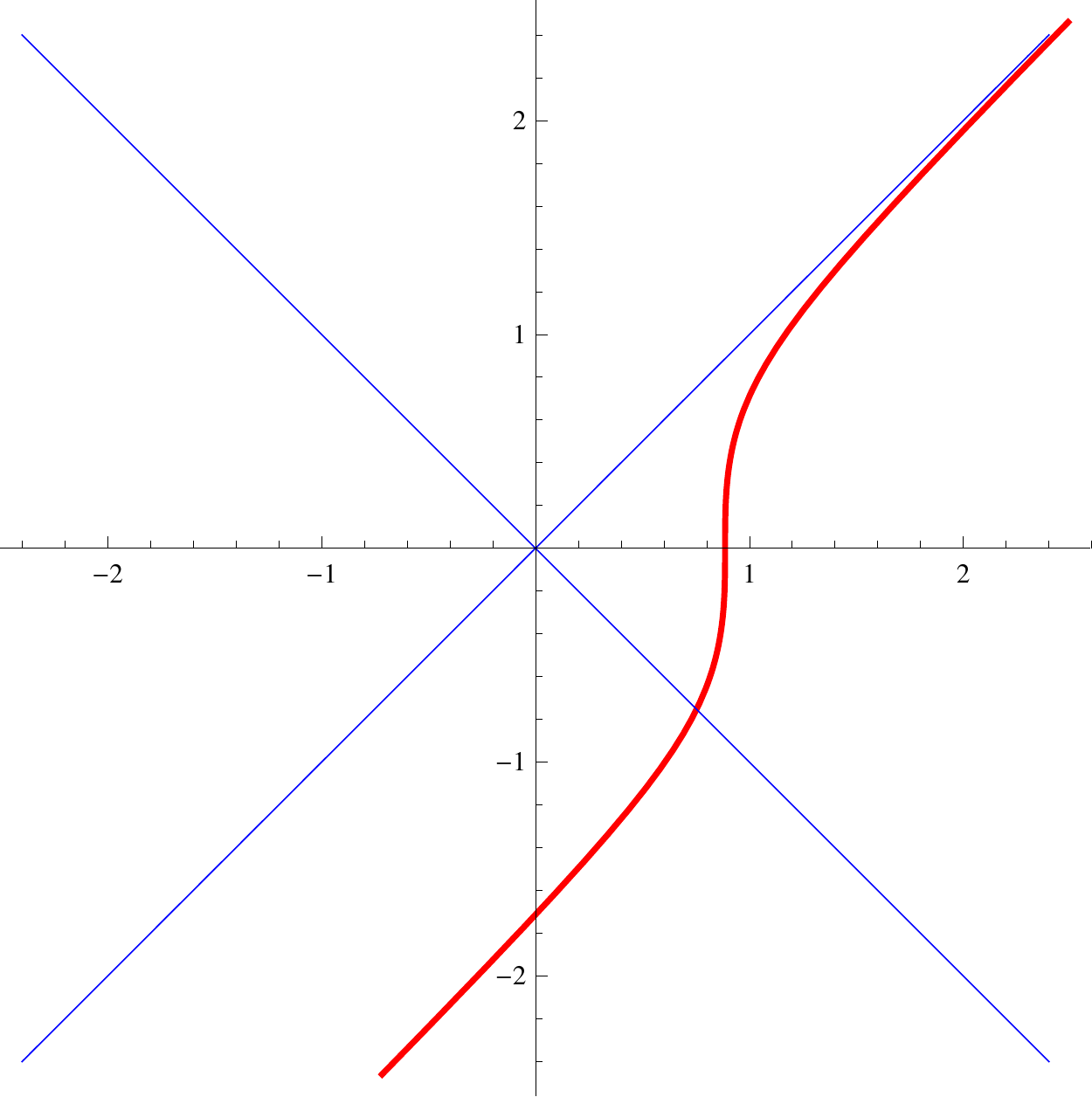}
\end{center}\caption{\sl 2d Minkowski. $n=2$, $\displaystyle P_2(s) = s^2$. 
The trajectory of the observer with vanishing Snap $=$ constant proper Jerk.  
Again, the line $x=t$ of the light cone is a horizon for the observer. Notice that, as it will happen
with all $n=$ even cases, the trajectory goes from $-\infty$ in the past
to $+\infty$ in the future.} 
\label{F2}
\end{figure}

\begin{figure}\begin{center}
\includegraphics[width=6cm,angle=0]{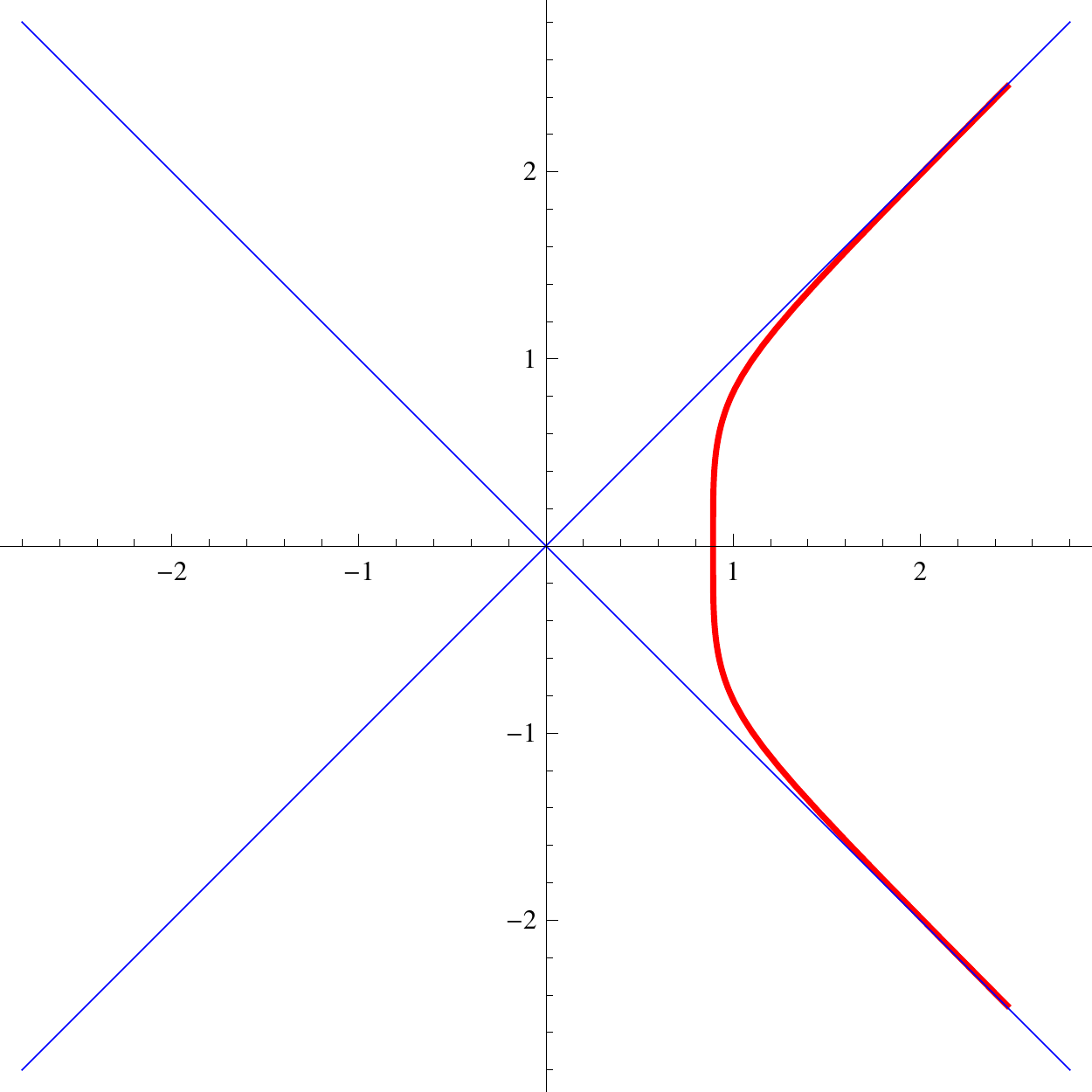}
\end{center}\caption{\sl 2d Minkowski. $n=3$, $\displaystyle P_3(s) = s^3$. 
The trajectory of the observer with vanishing Crackle $=$ constant proper Snap.} 
\label{F3}
\end{figure}

Here we depict
trajectories for $n=1$, constant proper acceleration (Fig \ref{F1}); $n=2$, constant proper Jerk 
(Fig \ref{F2}); and 
$n=3$, constant proper Snap (Fig \ref{F3}). 

\begin{figure}\begin{center}
\includegraphics[width=6cm,angle=0]{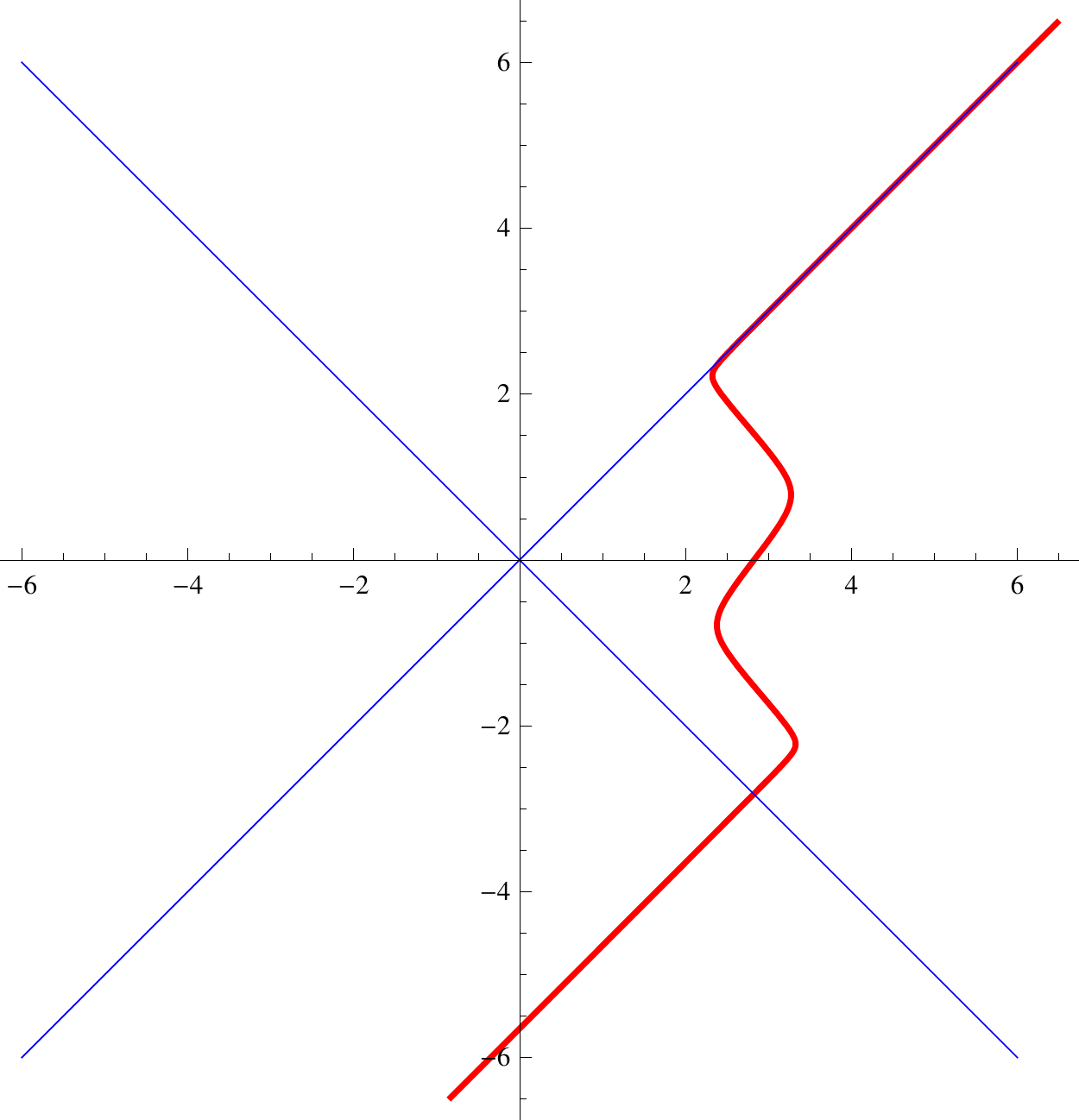}
\end{center}\caption{\sl 2d Minkowski. $n=4$, $\displaystyle P_4(s) = s^4-3s^2+1$. 
Now the trajectory of the observer with vanishing Pop $=$ constant proper Crackle
 exhibits four turning points, corresponding to the four real roots of $P_4(s)$.} 
\label{F4}
\end{figure}

In general one may consider, for an arbitrary polynomial $P_n(s)$ with real coefficients, 
its real roots, $s_k$. These roots are associated with the points in the trajectory 
$(X_0(s_k),\, \bar X_1(s_k))$  with vanishing $3$-velocity. If the root is simple or its degeneracly is 
an odd number, then 
the $3$-velocity changes sign at this point, and graphically we find a turning point in the trajectory, 
like the four turning points depicted in 
Fig (\ref{F4})\,.


\section{Conclusions}
\label{concl}

Starting within the context of Minkoswki spacetime and later generalizing to curved spacetime, 
we characterize the concept of a trajectory with constant proper acceleration by moving to the 
instantaneous comoving frame of reference of the observer, and choosing the boost that sends 
the observer to rest 
as a pure boost, with no additional rotation attached. We find that the motion of such an 
observer is at most three dimensional -including the time coordinate. Our characterization is 
formulated in the following way: "{\sl An observer undergoes a motion with constant proper acceleration 
if there exist an intertial frame of reference such that, using it for the description of the 
trajectory, the pure boost that sends the observer to rest is also sending the acceleration vector 
to a constant vector}". Formulated in such a way, this sentence does not seem to be exportable to 
curved spacetime. At this point the Frenet Serret formalism comes to the rescue and we show that 
our definition is equivalent to the statement that the third curvature invariant $\kappa_3$ 
vanishes ($\Rightarrow$ 
3d motion in Minkoswki spacetime), that the first and second curvature invariants $\kappa_1$
and $\kappa_2$ are constant and that $\kappa_2<\kappa_1$. In fact when $\kappa_2$ vanishes 
the motion is 2d and it is the standard textbook motion of the constantly accelerated observer 
in Minkoswki spacetime. This invariant characterization is immediately exported to curved spacetime as 
the definition of the observer with constant proper acceleration. Our definition excludes 
the uniform circular motion (see Appendix \ref{A}), for which $\kappa_2>\kappa_1$, with both 
$\kappa_2$ and $\kappa_1$ constants. 

\vs

Let us comment that our definition is much more restrictive than that used in \cite{Friedman:2016tye}, 
inspired in \cite{FR17}. In \cite{Friedman:2016tye} the observer with constant proper acceleration 
(uniform acceleration) is defined by the constancy
of the first, second and third curvature invariants (in a 4 dimensional Minkoswki spacetime), and 
it is understood that in higher dimensions the definition will include the constancy of the new 
curvature invariants that may appear. Of course it is all a matter of definition, but we would 
like to point out that with our definition, for whatever dimension of the Minkowski spacetime, 
there is always a frame of reference in which the motion will at most be three dimensional ($\kappa_3=0$).

We could still be more restrictive and argue that a reasonable definition of
constant proper acceleration should also include the vanishing of the Jerk. In such case the 
observer with constant proper acceleration would be characterized by $\kappa_2=0$ and 
$\kappa_1=\ constant$\,. As a matter of fact, the example of circular orbits in Schwarzschild spcetime 
(see Appendix \ref{C})) favors the adoption of such restrictive proposal for the definition 
of the constantly accelerated observer. 

Summing up, we may contemplate three possible characterizations of the constantly accelerated observer 
in curved spacetime, from the least to the most restrictive. 

a) All curvature invariants are constant.

b) $\kappa_3=0$ and $\kappa_1, \,\kappa_2 $ are constant with $\kappa_2<\kappa_1$ 
(our definition in section \ref{revisit}).

c) $\kappa_2=0$ and $\kappa_1$ is constant ($\Leftrightarrow$ vanishing Jerk).

Whereas the second case b) is supported by our procedure in section \ref{accobs}, 
we think that the case c) may be the most physically compelling. To implement this case in our formalism 
we should simply require the pure boosts that send the observer to rest at different times, 
to commute among themselves.  

\vs

We have studied the generalization of this concept of the accelerated observer 
(that is, with constant proper $1$-acceleration) to higher order time derivatives in Minkowski spacetime. 
All these new concepts allow for immediate extension to curved spacetime. The notions of vanishing 
$n$-acceleration and of constant proper $n$-acceleration are discussed
in detail for the cases of Jerk and Snap. We prove that the vanishing of the Jerk implies that the
proper acceleration is constant and in addition that the trajectory lies in a plane (2d)
in Minkowski spacetime. In contrast, we also show that the vanishing of the Snap does not imply that
the proper Jerk is constant, except if we restrict the motion to a plane (2d).

\vs

Although we have proved it only for $n=1\,, n=2$, we think it very likely that 
with higher order objects the notions of vanishing $(n+1)$-acceleration and of 
constant proper $n$-acceleration are equivalent only when the trajectory 
in Minkowski spacetime lies in a plane (or when $\kappa_2=0$ if we consider also curved spacetime). 
In general, it is not difficult to see that the notion of 
vanishing $n$-acceleration
has a specific characterization in the framework of the Frenet Serret formalism. 
We believe that the notion of constant proper $n$-acceleration, has also its 
characterization in such formalism, although we have not proved it beyond the case
of the constant proper acceleration. 
This would mean that all our results would be exportable to curved spacetime. 
We leave these issues for future work.

\vs

In Minkowski spacetime, when we specialize our results to trajectories in a plane, 
the characterization of vanishing $(n+1)$-acceleration together with 
constant proper $n$-acceleration is very simple 
by making use of the Frenet-Serret formalism. It all boils down to 
require that the second curvature invariant $\kappa_2$ vanishes and that the first curvature 
invariant $\kappa_1$ be (the absolute value of) a polynomial of degree $(n-1)$ of the
proper time parameter, This result can be exported to generally covariant metric theories with 
a connection satisfying the metricity condition.

\vs

Concerning motion in a plane, we may compare our results  with those in Galilean mechanics. 
In Galilean 
mechanics with one space dimension, the velocity of an observer with constant proper $n$-acceleration
is a $n$-degree polynomial $P_n(t)$ of the physical time $t$. 
In Minkowski 2d spacetime the velocity of an observer with constant proper 
$n$-acceleration is $v(t)= \tanh(P_n(s(t)))$ where $P_n$ is an $n$-degree polynomial and 
$s(t)$ -proper time as function of coordinate time- is obtained after solving
the differential equation $\displaystyle \frac{d\,t}{d\,s} = \cosh(P_n(s))$. Thus moving form 
Galilean to Minkowskian observers amounts to encapsulate the Galilean results within specific
hyperbolic trigonometric functions that set up the bound of the speed of light. 

\section{Acknowledgements}

JMP thanks Joaqu\'{\i}m Gomis and Carles Batlle for useful conversations. He 
also acknowledges financial support from  projects 2017-SGR-929, MINECO grant FPA2016-76005-C2-1-P and the 
MDM-2014-0369 of ICCUB (Unidad de Excelencia ‘Mar\'{\i}a de Maeztu’).

\begin{appendix}


\section{Uniform circular motion}
\label{A}

Consider in a 3d Minkowski spacetime the trajectory ($r>0 $ and $\omega> 0$ are constant)
\beq
X(t(s))=\left(
\begin{array}{c}
t(s)\\ r \sin(\omega\, t(s))\\ -r \cos(\omega\, t(s))\\
\end{array}
\right) \,.
\label{circ3d}
\eeq
Its velocity with respect to proper time $s$ is
$$
V(s)=\dot t(s)\left(
\begin{array}{c}
1\\ r\,\omega \cos(\omega\, t(s))\\ r\omega \sin(\omega\, t(s))\\
\end{array}
\right) \,,
$$
and the condition $(V)^2 = -1$ determines 
$\displaystyle\dot t(s) = \frac{1}{\sqrt{1-r^2 \omega^2}} \equiv \gamma$\,. 
Note that $r,\ \omega$ are required to satisfy  $r\, \omega < 1$.

\vs

The curvature invariant $\kappa_3$ vanishes -the motion is 3d- and we obtain 
$\displaystyle \kappa_1= \frac{r \omega^2}{1-r^2 \omega^2}\,,\ 
\kappa_2 =  \frac{\omega}{1-r^2 \omega^2}$\,. Thus $\kappa_1 = r\,\omega\,\kappa_2 < \kappa_2$. 
Adopting the procedure in section \ref{revisit} we can easily recover from the constancy 
of $\kappa_1\, (< \kappa_2)$ and  $\kappa_2$ the solution  (\ref{circ3d}). 

\vs

One can systematically compute the $n$-accelerations. 
Let us define the unitary, spacelike vectors  $T$  and  $N$ 
(the spatial component of $T$ is {\sl tangent} to the spatial component of $V$ and 
the spatial component of $N$ is {\sl normal} to the spatial component of $V$),
$$
T=\gamma\left(
\begin{array}{c}
r\,\omega\\ \cos(\omega\, t(s))\\ 
\sin(\omega\, t(s))\\
\end{array}
\right)  \,,\qquad N=\left(
\begin{array}{c}
0\\ -\sin(\omega\, t(s))\\ 
\cos(\omega\, t(s))\\
\end{array}
\right)
$$
So we have $T^2=N^2=1$,\ $(T,N)=(T,V)=(N,V)=0$.
It is then a matter of straightforward computation to see that the $n$-accelerations $A_n$ can be 
written as
$$
A_{2k+1}= (-)^k\, r\,\omega (\gamma^2\omega )^{2k+1} N,\ k=0,1,2...
$$
$$
A_{2k}= (-)^k\, r\,\omega (\gamma^2\omega )^{2k} T,\ k=1,2,3...
$$
Thus in this simple case of uniform circular motion all the $n$-accelerations are non-vanishing, 
though the norm of each one is constant.

The proper $n$-acceleration is obtained by applying to $A_n$ the pure boost (\ref{theboost2}). 
Noticing that $B\,N = N$ and $B\, T= \gamma (T-r\omega V)$ we observe that all the proper
$n$-accelerations $A_n^{[p]}$ exhibit a dependence of periodic type on proper time. 

\section{Constantly accelerated observers in an expanding de Sitter universe}
\label{B}
We consider the de Sitter metric with flat equal-time slices,
\beq
ds^2 = -dt^2 + ({\rm e}^{\sqrt{\frac{\Lambda}{3}}\,t})^2\Big(dx^2 + 
dy^2 + dz^2\Big)\,,
\label{dSmetric}
\eeq
where $\Lambda$ is the cosmological constant, $\Lambda>0$. With this parametrization, 
the comoving observers ($\vec x = $ {\sl constant}) are geodesics.

We will show that the timelike trajectory 
\beq
x(t) = (x_0-u) + u \,{\rm e}^{-\sqrt{\frac{\Lambda}{3}}\,t}\,,\ y=y_0\,,\ z=z_0\,,
\label{accdesitter}
\eeq
(the constant parameter $u$ will be restricted below) 
satisfying $x(0)=x_0$, exhibits  vanishing Jerk, which implies constant proper acceleration.

In fact, the velocity with respect to proper time ($(V)^2 = V^\mu g_{\mu\nu} V^\nu =-1$) is
$$
V=\frac{1}{\sqrt{1-\frac{u^2\Lambda}{3}}}\left(
\begin{array}{c}
1 \\-u\,\sqrt{\frac{\Lambda}{3}}{\rm e}^{-\sqrt{\frac{\Lambda}{3}}\,t}
 \\ 
 0\\
 0
\end{array}
\right)\,.
$$
Note that we need $|u|< \sqrt{\frac{3}{\Lambda}}$, otherwise the trajectory becomes lightlike (for 
$ |u|= \sqrt{\frac{3}{\Lambda}}$) or spacelike (for 
$ |u|> \sqrt{\frac{3}{\Lambda}}$).

Note that the coordinate time is proportional to the proper time, 
$\displaystyle \frac{d\,t}{d\,s}= \frac{1}{\sqrt{1-\frac{u^2\Lambda}{3}}}\,$.
The acceleration becomes
$$
A^\mu = \frac{d}{d\,s} V^\mu + V^\rho\Gamma_{\rho\nu}^\mu V^\nu 
= \frac{1}{\sqrt{1-\frac{u^2\Lambda}{3}}}\frac{d}{d\,t} V^\mu + V^\rho\Gamma_{\rho\nu}^\mu V^\nu 
= \left(
\begin{array}{c}
\frac{u^2\Lambda ^{3/2}}{\sqrt{3} \left(3-u^2\Lambda\right)} \\ 
-\frac{u\,\Lambda}{3-u^2\Lambda}{\rm e}^{-\sqrt{\frac{\Lambda}{3}}\,t}
 \\ 
 0\\
 0
\end{array}
\right)\,,
$$
which satisfies
$$
(A)^2 = A^\mu g_{\mu\nu}A^\nu= \frac{u^2\Lambda^2}{9-3\, u^2\Lambda}\,.
$$
Note that for $u=0$ the motion is geodesic. Finally, we check that the Jerk vanishes,
$$
\Sigma^\mu= \frac{d}{d\,s} A^\mu + V^\rho\Gamma_{\rho\nu}^\mu A^\nu - (A)^2 V^\mu =0\,.
$$
One can compute also curvature invariants. We obtain
$\displaystyle \kappa_1= \frac{u \,\Lambda}{\sqrt{9-3\, u^2\Lambda}}$\,, $\kappa_2=0$. We may notice that,
in view of (\ref{jerk-FS}), which is valid for our curved spacetime, 
the constancy of $\kappa_1$ and the vanishing of $\kappa_2$ 
already guarantee the vanishing of the Jerk.
\subsection{\bf Horizon for the accelerated observer}
Now we compare the trajectory of our accelerated observer with that of light. In the $x$ direction
the two lightlike trajectories with initial condition $\bar x(0) =  \bar x_0$ are 
\beq
\bar x(t) =  (\bar x_0 \mp \sqrt{\frac{3}{\Lambda}}) 
\pm \sqrt{\frac{3}{\Lambda}}{\rm e}^{-\sqrt{\frac{\Lambda}{3}}\,t}\,,\ \bar y=y_0\,,\ \bar z=z_0\,,
\label{lighttr}
\eeq
so that $\bar x(0) =  \bar x_0$. 

Let us find conditions on $x_0$ and $\bar x_0$ to guarantee that at some moment
in the future or in the past both trajectories intersect. The geodesic line between the two 
trajectories taken at time $t_0$ is, for the $x$ coordinate ($\lambda\in [0,\,1]$)
$$
\hat x(\lambda) = \lambda\, x(t_0) +(1-\lambda)\,\bar x(t_0)\,,
$$
and $\hat y(\lambda) =0,\ \hat z(\lambda) =0$\,. The distance $d$ at $t_0$ between the two 
trajectories at this time $t_0$ is
$$
d\Big(\bar x(t_0),x(t_0)\Big) =\int_0^1 d\lambda \sqrt{\frac{d\hat x}{d \lambda}g_{xx} 
\frac{d\hat x}{d \lambda}} =
|(\bar x(t_0) - x(t_0)) {\rm e}^{\sqrt{\frac{\Lambda}{3}}\,t_0}|\,.
$$
For this distance to vanish at some finite time $t_0$ we need
$$
0= \Big|(\bar x(t_0) - x(t_0)){\rm e}^{\sqrt{\frac{\Lambda}{3}}\,t_0}\Big|= 
\Big|\Big(\bar x_0 - x_0 +(u \mp \sqrt{\frac{3}{\Lambda}})\Big)
{\rm e}^{\sqrt{\frac{\Lambda}{3}}\,t_0} - (u \mp \sqrt{\frac{3}{\Lambda}})\Big|\,
$$
and this equation has solution for $t_0$ if and only if 
$$
\frac{\bar x_0 - x_0 +(u \mp \sqrt{\frac{3}{\Lambda}})}{(u \mp \sqrt{\frac{3}{\Lambda}})}>0\,,
$$
which is equivalent to 
$$
\Big|\bar x_0 - (x_0-u)\Big|< \sqrt{\frac{3}{\Lambda}}.
$$
This is the condition for the accelerated observer and lightlike trajectories to intersect 
somewehere.
The non-intersection condition sets up a horizon -here at time $t=0$- located at $\bar x_0=x_H(0)$,
\beq
x_H(0) = (x_0-u)\pm \sqrt{\frac{3}{\Lambda}}.
\label{xhor}
\eeq
This horizon moves at the speed of light,
\beq
x_H(t) = (x_0-u)\pm \sqrt{\frac{3}{\Lambda}}{\rm e}^{-\sqrt{\frac{\Lambda}{3}}\,t}\,.
\label{xhort}
\eeq
The picture given by (\ref{xhort}) is incomplete and can be misleading since 
it seems that there are two horizons. In view of (\ref{accdesitter}) and (\ref{xhort}) it is
convenient to adopt spherical coordinates for the equal-time $3$-surfaces centered in 
$x= x_0-u,\ y=y_0,\ z=z_0$. 
With $r$ the radial coordinate, the trajectory of the accelerated observer is then
\beq
r(t)= u\,{\rm e}^{-\sqrt{\frac{\Lambda}{3}}\,t }\,,
\label{accrad}
\eeq
with the angular variables remaining fixed. Now the horizon appears in the equal-time surfaces as a 
sphere around $x_0-u,\ y=y_0,\ z=z_0$ with
coordinate radius
$$
r_H(t) =  \sqrt{\frac{3}{\Lambda}}{\rm e}^{-\sqrt{\frac{\Lambda}{3}}\,t}\,.
$$
moving at the speed of light. Notice that this is nothing but the future event horizon 
\cite{Hawking:1973uf}, also known as the cosmological horizon of de Sitter.

Note that for $t\to \infty$ both $r(t)$ and $r_H(t)$ tend to $r=0$, but this is an effect of the 
coordinatization. As a matter of fact all equal-time distances between the center $r=0$, 
the trajectory of the accelerated observer, and the horizon, remain constant. 
For instance the distance between the trajectory and the horizon is, 
for the same values of the angular variables,
$$
D\Big(r(t),r_H(t)\Big) = |(r(t)-r_H(t)){\rm e}^{\sqrt{\frac{\Lambda}{3}}\,t}|= 
\sqrt{\frac{3}{\Lambda}}-u\,.
$$
whereas the distance of the accelerated observer to the center is $u$ and the length
of the radius of the spherical horizon is $\sqrt{\frac{3}{\Lambda}}$.

\subsection{\bf Comoving observers of different kind}

In fact one can use (\ref{accrad}) to define the standard static coordinates for de Sitter. 
We know that the comoving observers for the metric (\ref{dSmetric}) are geodesics, and we also know that
the radial motion that keeps $r\,{\rm e}^{\sqrt{\frac{\Lambda}{3}}\,t}$ constant 
is that of a constantly accelerated observer.
If we define the new radial coordinate $u=r\,{\rm e}^{\sqrt{\frac{\Lambda}{3}}\,t}$ then 
this accelerated observer will sit at constant $u$. The metric becomes 
\bea
ds^2= -dt^2 + {\rm e}^{\sqrt{\frac{\Lambda}{3}}\,t}\Big(dr^2 + 
r^2(d\theta^2 + (\sin\theta)^2 d\varphi^2)\Big)\ \rightarrow \nonumber\\ 
-(1-\frac{\Lambda}{3}u^2) dt^2 + du^2  -2 u \sqrt{\frac{\Lambda}{3}} dt\, du
+ u^2(d\theta^2 + (\sin\theta)^2 d\varphi^2)\,.
\eea
Next, to get rid of the non-diagonal term, we solve a differential equation and obtain
the change of variables
$\displaystyle T = t-\frac{1}{2} \sqrt{\frac{3}{\Lambda}}
\log(1-\frac{\Lambda}{3}u^2)$. We get
\beq
ds^2=-(1-\frac{\Lambda}{3}u^2) dT^2 + \frac{1}{1-\frac{\Lambda}{3}u^2}du^2  
+ u^2(d\theta^2 + (\sin\theta)^2 d\varphi^2)\,.
\label{dStwo}
\eeq
So, in de Sitter spacetime, whereas the comoving observers in the coordinatization of (\ref{dSmetric}) 
are geodesics, the comoving observers in (\ref{dStwo}) are constantly accelerated observers. 

\vs

A similar construction can be performed for 
Anti de Sitter spacetime. The comoving observers in the coordinates of the metric 
(\ref{dStwo}), now with $\Lambda<0$, exhibit
constant proper acceleration. Looking for the radial geodesics in this form of the metric we can move
to a coordinatization in which such geodesics are the trajectories of the comoving observers. 
Unsurprisingly, one obtains the ``cosmological'' form of AdS metric ($a=\sqrt{-\frac{\Lambda}{3}}$),
\beq
ds^2=- dt^2 + (\cos(a\, t))^2\Big(\frac{1}{1+a^2 r^2}dr^2  
+ r^2(d\theta^2 + (\sin\theta)^2 d\varphi^2)\Big)\,.
\label{AdStwo}
\eeq

\section{Circular orbits in Schwarzschild spcetime}
\label{C}

With coordinates $(t,r,\theta,\varphi)$, we consider Schwarzschild metric
\beq
ds^2=- (1-\frac{2\, M}{r}) dt^2 + \frac{1}{(1-\frac{2\, M}{r})}dr^2  
+ r^2(d\theta^2 + (\sin\theta)^2 d\varphi^2)\Big)\,,
\label{schw}
\eeq
and circular uniform timelike trajectories $(t,r_0,\frac{\pi}{2},\omega\, t)$ 
outside the event horizon, $r_0>2\,M$. In terms of proper time $s$, the velocity vector is
$V = \dot t(s) (1,0,0,\omega)$ and the requirement $V^2=-1$ yields 
$\displaystyle  \dot t = (1-\frac{2\, M}{r_0}-r_0^2\omega^2)^{-\frac{1}{2}}$ which sets the bound 
for $\omega$, 
\beq
\displaystyle \omega^2 < \frac{r_0-2\, M}{r_0^3}\,,
\label{omegabound}
\eeq
to keep the  trajectory timelike.
Using the definitions in section \ref{FS}, extended through section \ref{Extension} 
to curved spacetime, it is easy to compute the curvature scalars. We get
\beq
\kappa_1= \frac{\sqrt{r_0-2 M} \left| M-r_0^3\, \omega^2\right|
}
{r_0^{\frac{3}{2}}(r_0-2\, M -r_0^3\, \omega^2)}\,,\qquad 
\kappa_2= \frac{\omega  \left| r_0-3 M\right| }{(r_0-2\, M -r_0^3\, \omega^2)}\,,\qquad \kappa_3=0\,.
\eeq
The condition $\kappa_1= 0$, that is, $r_0^3\, \omega ^2=M$, identifies geodesic motion for our circular 
orbits. In this case, since $\omega$ already satifies (\ref{omegabound}) we end up with the well 
known condition $r_0>3 \,M$ for the existence of timelike geodesics in circular motion 
(the circular geodesic becomes lightlike for $r_0=3\,M$). Now we have the aditional information 
that the second curvature invariant vanishes for circular orbits at $r_0=3\,M$. For these orbits at 
$r_0=3\,M$ to be timelike it suffices to require $\displaystyle \omega ^2< \frac{M}{r_0^3}$. 
Thus these circular orbits, with constant $\kappa_1$ and vanishing $\kappa_2$ correspond to our most 
restrictive version of constantly accelerated observers.

Let us turn to the less restrictive version considered in sections \ref{the3dsol},\,\ref{revisit}, 
so that we require $\kappa_1$ and $\kappa_2$ to be constant but with $\displaystyle \kappa_2<\kappa_1$. 
Let us compute the quantity (with $X := r_0^3\, \omega ^2 < r_0-2\, M$)
$$
Q=\Big(\frac{\kappa_2}{\kappa_1}\Big)^2 = \frac{X (r_0-3\,M)^2}{(r_0-2\, M) (M-X)^2}\,,
$$
and ask for values of $X$, compatible with $X < r_0-2\, M$ , such that $Q<1$. We find
\bea
{\bf a)}&& \  {\rm for} \ r_0 > 3\,M \,, \ {\rm any}\ X \ {\rm chosen\ as} \ X < \frac{M^2}{r_0-2\, M}\,,
\nonumber\\
{\bf b)}&& \  {\rm for} \ 2\,M < r_0 \leq 3\,M \,, \ {\rm any}\ X \ {\rm chosen\ as} \ X <r_0-2\, M\,.
\nonumber
\eea
Note that in both cases $X<M$ (recall that $X=M$ is geodesic motion, which in fact is 
impossible for time like trajectories in the case {\bf b)}).
The {\bf b)} case is quite understanable: $\omega$ is allowed to have any value compatible with the 
speed of light bound $X <r_0-2\, M $. Instead, the requirement 
$\displaystyle X < \frac{M^2}{r_0-2\, M} $ for the case {\bf a)} could make us think that something
special happens at the ``critical'' value $\displaystyle X=\frac{M^2}{r_0-2\, M} $; 
we do not see anything special at this value for $X$, which makes $\displaystyle \kappa_2=\kappa_1= 
\frac{M \sqrt{r_0-2 M}}{r_0^{3/2} (r_0-M)}$. Thus if this ``critical'' value has no meaning other than that, 
it would seem more natural to adopt the most restrictive version of the notion of the 
constantly accelerated observer, that is, the vanishing of $\kappa_2$ and constancy of $\kappa_1$. This 
singles out, in the case of circular orbits, those at $r_0 = 3\,M$.

\end{appendix}

\end{document}